\documentclass[a4paper,11pt]{article}

\usepackage{graphicx}
\usepackage{pgfplots,pgfplotstable}
\pgfplotsset{compat=1.18}
\usepgfplotslibrary{groupplots,fillbetween}

\usepackage[utf8]{inputenc}
\usepackage[T1]{fontenc}

\usepackage[DIV=15]{typearea}

\usepackage{microtype}
\usepackage{lmodern}
\usepackage{xspace}
\usepackage{cite}
\usepackage{paralist}
\newcommand{\eps}{\varepsilon}

\usepackage{amsmath}
\usepackage{amssymb}
\usepackage{amsthm}
\usepackage{mathtools}

\usepackage{tikz}
\usetikzlibrary{arrows.meta,positioning,decorations.pathreplacing}

\usepackage{hyperref}
\usepackage[capitalise,noabbrev]{cleveref}

\theoremstyle{plain}
\newtheorem{theorem}{Theorem}
\newtheorem{proposition}[theorem]{Proposition}
\newtheorem{lemma}[theorem]{Lemma}
\newtheorem{corollary}[theorem]{Corollary}

\newtheorem{observation}[theorem]{Observation}
\theoremstyle{definition}

\newtheorem{example}[theorem]{Example}

\title{Ordinal Hegselmann-Krause Systems}
\date{}
\author{Martin Hoefer \thanks{Faculty of Computer Science, RWTH Aachen University, Germany. \texttt{mhoefer@cs.rwth-aachen.de}}
\and Tim Koglin \thanks{Faculty of Computer Science, RWTH Aachen University, Germany. \texttt{koglin@algo.rwth-aachen.de}}
\and Jakob Lindner \thanks{Faculty of Computer Science, RWTH Aachen University, Germany. \texttt{lindner@algo.rwth-aachen.de}}
\and Jannis Schramek \thanks{Faculty of Computer Science, RWTH Aachen University, Germany. \texttt{jannis.schramek@rwth-aachen.de}}
}

\newcommand{\notation}[1]{\ensuremath{#1}\xspace}

\newcommand{\agg}{\notation{\textsf{\upshape{agg}}}}
\newcommand{\minagg}{\notation{\textsf{min}}}
\newcommand{\maxagg}{\notation{\textsf{max}}}
\newcommand{\medianagg}{\notation{\textsf{median}}}
\newcommand{\largeagg}[1]{\notation{{#1}\textsf{-large}}}
\newcommand{\smallagg}[1]{\notation{{#1}\textsf{-small}}}

\newcommand{\nopns}[2]{\notation{X_{\mathcal{N}_{#2}}(#1)}}
\newcommand{\nh}[2]{\notation{\mathcal{N}_{#2}(#1)}}

\renewcommand{\O}{\notation{\mathcal{O}}}

\def\abs#1{\lvert#1\rvert}

\newcommand{\opn}[2]{\notation{x_{#2}(#1)}}
\newcommand{\topn}[2]{\notation{\tilde{x}_{#2}(#1)}}
\newcommand{\opns}[1]{\notation{\boldsymbol{x}_{#1}}}
\newcommand{\topns}[1]{\notation{\tilde{\boldsymbol{x}}_{#1}}}
\newcommand{\dcto}[1]{\notation{\ell_{#1}}}

\newcommand{\disopn}[1]{\notation{\chi_{#1}}}
\newcommand{\onh}[1]{\notation{N(#1)}}
\newcommand{\loads}[1]{\notation{n^>_{#1}}}
\newcommand{\load}[2]{\notation{\loads{#2}(#1)}}

\newcommand{\ipot}[2]{\notation{\Phi_{#1}(\opns{#2})}}
\newcommand{\mov}[1]{\ensuremath{\textnormal{mov}_{#1}^t}\xspace}

\newcommand{\lmulti}{{\{\hspace{-3.4pt}\{}}
\newcommand{\rmulti}{{\}\hspace{-3.4pt}\}}}

\def\N{\notation{\mathbb{N}}}
\def\R{\notation{\mathbb{R}}}

\begin{document}
\maketitle

\begin{abstract}
We study opinion formation in the popular Hegselmann-Krause model when agents replace the traditional averaging update by a median update or more general ordinal aggregation methods. We analyze the convergence properties of synchronous and asynchronous dynamics.

For asynchronous dynamics, we show that every HKS with a $1$-Lipschitz ordinal aggregation function converges in worst-case time $\O(n^4)$. For the special case of median aggregation and for the family of \largeagg{k} aggregation functions, in which agents select the $k$-th largest opinion in their neighborhood, we obtain sharper bounds of $\O(n^3)$. For general ordinal aggregation, some sequence of $\O(n^2)$ updates reaches an equilibrium from every initial state. Nevertheless, other update sequences may cycle, and uniform random activation can require exponentially many updates with overwhelming probability.

For synchronous dynamics, we obtain a tight characterization. Ordinal aggregation functions satisfying the $1$-Lipschitz condition are shown to guarantee convergence in linear time. This result applies to many natural examples such as minimum, maximum, median, and \largeagg{k} functions. When the Lipschitz condition is violated, there exist instances in which synchronous dynamics do not converge at all.
\end{abstract}

\section{Introduction}

The course of modern societies is substantially driven by the opinion formation of their inhabitants. Opinion formation plays a major role in the outcome of political discussions, the economic success of products and companies, or in leading to new trends and directions in research and development. Understanding how opinions are formed is essential for managing conflicts and designing effective communication strategies. Consequently, there has been a large research interest in modeling and analyzing the formation of opinions across different fields, including AI, sociology, control theory, physics, and many more.

Classic opinion formation models in the literature involve a social network as a fixed environment, in which agents adjust their opinions according to peer pressure and intrinsic beliefs (such as French-DeGroot~\cite{French1956,Degroot1974} or Friedkin-Johnsen~\cite{FriedkinJohnsen}). Since the early 2000s, research focus has shifted towards \emph{bounded confidence} models, in which each agent limits attention to other agents with similar opinions in a region around its current opinion. Arguably, the most prominent of these bounded-confidence models is the Hegselmann-Krause system (HKS)~\cite{HegselmannKrause}. There are $n$ agents, and at time $t \ge 1$, each agent $i$ holds a current opinion $x_t(i)$. Each agent $i$ considers the set $\nh{i}{t}$ of all agents $j$ in a bounded neighborhood, i.e., for which $|\opn{j}{t} - \opn{i}{t}| \le \varepsilon$, for a given constant $\varepsilon > 0$ (where, in particular, $i \in \nh{i}{t}$). For time $t+1$, they update their opinion so that it matches the \emph{average} of these opinions in the bounded neighborhood, i.e., $x_{t+1}(i) = \frac{1}{|\nh{i}{t}|} \sum_{j \in \nh{i}{t}} x_t(j)$.

The averaging behavior of agents is a predominant property in a large number of opinion formation models. This behavior can be interpreted in a game-theoretic sense by assuming each agent strategically chooses an opinion in order to minimize a 2-norm of the vector of distances to neighboring opinions~\cite{BindelKO15}. While being extremely popular in formal models, averaging behavior is often unrealistic in practice. As a consequence, there is a recently emerging interest in other aggregation methods. In particular, several works have started to explore opinion formation based on the \emph{median} rather than the average~\cite{Mei2022,MeiTAC24,ZhangTAC25,ZhangSCIS2025,BerenbrinkAIJ2026,mohr2025} since it has a solid underpinning in cognitive psychology and entails favorable statistical properties. Further, aggregation via the median has desirable properties in social choice~\cite{Moulin80,Ching97}. Note that the median can be interpreted as the result of agents strategically minimizing the 1-norm of the vector of distances to neighboring opinions.

\paragraph{Contribution and Outline.}
In this paper, we explore the impact of median and more general ordinal aggregation rules on opinion formation with bounded confidence in HKS. In an ordinal aggregation function, each agent selects their most preferred opinion from the sorted order of opinions in their neighborhood. This process is determined only by the current size and ordering of that neighborhood and regardless of the exact numerical opinion values. Since agents only adopt opinions held by other agents, the set of opinions held in every step is a subset of the set of initial opinions. %

For HKS with such an ordinal aggregation of agents (which we call ordinal HKS), we examine the conditions and the time needed for the system to converge to an equilibrium. In an equilibrium, all agents are stable, i.e., the current opinion of each agent equals the aggregated opinion of their current neighborhood.
Our results inform the discussion of the predictive value of these models, i.e., whether or not it is realistic to expect that a stable opinion configuration emerges from ordinal aggregation (in reasonable time). We study convergence properties of both synchronous and asynchronous dynamics. For synchronous dynamics, all agents that are not stable update their opinion simultaneously at each time step; for asynchronous dynamics, exactly one agent that is not stable is chosen at each time step to update their opinion.
We stress that convergence times in the two models are measured on different scales. A synchronous step may comprise between $1$ and $n$ simultaneous individual updates, whereas an asynchronous step comprises exactly one. Consequently, bounds on the number of synchronous and asynchronous steps should not be compared directly.
To the best of our knowledge, there is no prior work on the impact of alternative aggregation functions in HKS that provides rigorous quantitative bounds on the convergence behavior of such scenarios.

After a formal introduction of the model and a discussion of ordinal aggregation functions in \cref{sec:model}, we discuss our results on asynchronous dynamics in \cref{sec:async}. We show that every ordinal HKS is weakly-acyclic and best-case convergence time is polynomial, i.e., for every initial configuration there is a sequence of $\O(n\ell)$ moves that leads the system to an equilibrium. This implies that the system converges in the limit with probability 1 when the moving agent is selected uniformly at random at each time step. On the other hand, we give a class of instances that can lead to cyclic behavior for uniform random activation. We also show that the average-case convergence time can be exponential in $n$ with overwhelming probability.

Towards worst-case convergence time, we show polynomial bounds for a number of important classes of aggregation functions. For median aggregation, we obtain a bound of $\O(n^2\ell)$. Beyond the median, we explore the natural classes of \largeagg{k} and \smallagg{k} aggregation, in which every agent adopts the $\min\{k,|\nh{i}{t}|\}$-largest or $\min\{k,|\nh{i}{t}|\}$-smallest opinion from $\nh{i}{t}$. These aggregation functions result in a selection via $k$-th order statistics, for a fixed value of $k$. We focus our discussion on \largeagg{k} aggregation throughout, but all our results apply symmetrically to \smallagg{k} aggregation.

When the agents always prefer the largest ($k = 1$) or smallest ($k = n$) opinion in their neighborhood, the worst-case convergence time is $\O(n \ell)$. For general $k$, we construct an integer-valued potential and prove that every asynchronous \largeagg{k}-HKS converges in worst-case time $\O(n \ell^2)$. By symmetry, the same guarantee holds for \smallagg{k} aggregation.

As a broad extension of median and \largeagg{k} aggregation, we show that every HKS with a $1$-Lipschitz ordinal aggregation function converges to an equilibrium in (worst-case) time $\O(n^2\ell^2)$. This is tight in the following sense. There is a strictly $2$-Lipschitz aggregation function for which there is a HKS with cyclic dynamics and, thus, infinite worst-case convergence time. More generally, \emph{every} strictly $k$-Lipschitz aggregator with $k \ge 3$ allows us to construct such a HKS.

For synchronous dynamics, we show in \cref{sec:sync} that the order-preserving ordinal aggregation functions are exactly the non-decreasing functions whose rank selector increases by at most $1$ when the neighborhood size increases by $1$. We call these functions $1$-Lipschitz. They guarantee convergence in at most $n-1$ steps and include minimum, maximum, median, and \largeagg{k} aggregation. This is again tight in the Lipschitz hierarchy. There is a strictly $2$-Lipschitz aggregation function for which there is a HKS where convergence fails.

\paragraph{Related Work.}
The standard averaging-based HKS has attracted substantial research interest, primarily driven by experimental results and simulations.
In contrast, the literature on analytical results and provable structural and computational worst-case bounds is comparatively sparse. We concentrate our discussion of related work on analytic convergence results and provable bounds on the convergence time in variants of HKS.

The convergence properties of synchronous dynamics in the standard HKS are rather well-understood, even for higher-dimensional opinion spaces. For synchronous dynamics on complete networks, it is known that the process always converges to a stable state, even for generalizations with symmetric and time-dependent visibility~\cite{Lorenz2005}, stubborn agents~\cite{ChazelleW17}, or when agents only consider a random subset of neighbors in each step~\cite{FotakisPS16}.
The convergence time in the one-dimensional HKS is upper bounded by $\O(n^3)$~\cite{BhattacharyyaITCS13}. The latter work also establishes a lower bound of $\Omega(n^2)$ on the convergence time, which holds even in one dimension~\cite{Wedin2015}.
For higher-dimensional opinions, the first upper bound was $\O(n^{10} d^2)$~\cite{BhattacharyyaITCS13}, which was subsequently improved to $\O(n^6)$~\cite{EtesamiBasar15} and further decreased to $\O(n^4)$~\cite{Martinsson16}.
Convergence times have also been studied when agents are apriori restricted in their interaction by a social network. Here each agent further restricts the opinions in the bounded confidence interval to those held by neighbor agents in the network. For such HKS with arbitrary social networks, significantly weaker guarantees are known. A synchronous HKS reaches a $\delta$-stable state (in which all opinions are either at distance at most $\delta$ or at least $\varepsilon$) within $\O(n^5(\varepsilon/\delta)^2)$ rounds~\cite{BhattacharyyaCDC15}. In contrast, the expected convergence time to a perfectly stable state can be infinite due to persistent oscillations~\cite{ParasnisTAC22}.

For asynchronous dynamics, finite-time convergence is not guaranteed. Instead, convergence to $\delta$-stable states (also termed $\delta$-equilibria) has been studied and bounded by $\O(n^9 (\varepsilon/\delta)^2)$~\cite{EtesamiBasar15}. The average-case convergence time of asynchronous dynamics was subsequently improved to $\O(n^5 + n^3(\varepsilon/\delta)^2)$, and there is a lower bound of $\Omega(n^4)$ when $\varepsilon/\delta$ is constant~\cite{BerenbrinkDC2024}. The upper bound can be generalized to $O(n m^2 (\varepsilon/\delta)^2)$ for HKS on arbitrary social networks with $m$ edges. These bounds apply when the random agent selected for update can be one that currently has no incentive to deviate.

Opinion formation processes are studied intensively in AI and multi-agent systems. In particular, our paper relates to a body of work that analyzes opinion formation models from a game-theoretic perspective~\cite{BhawalkarSTOC13,BindelKO15,BiloTCS18,EpitropouSAGT17}. This literature typically focuses on the existence of equilibria and their social quality, measured by the price of anarchy. More closely related to our work are works on equilibrium existence and convergence in discrete preference games~\cite{Chierichetti2018, FerraioliTCS16, AulettaAAMAS17}. Here each agent has a fixed neighborhood in a social network as well as a static intrinsic belief (similar to the Friedkin-Johnsen model~\cite{FriedkinJohnsen}). In contrast to HKS, discrete preference games are usually not bounded-confidence models, in which the neighborhood changes based on proximity of chosen opinions. Instead, there is a metric space for measuring the distance of any pair of neighboring agents based on their adopted opinions, and agents pick opinions to minimize the total (or average) distance to their fixed neighbors. Due to symmetry in distance, these games have exact potential functions, and convergence of asynchronous better-response dynamics is guaranteed. Existence of ordinal potential functions can also be observed in related variants with binary opinions and additional constraints on the agent preferences~\cite{AulettaCFGP16}. The computation of equilibria can be \textsf{PLS}-hard~\cite{lolakapuri2019}, and the best-case convergence time of asynchronous dynamics can be exponential. In some game classes, equilibria can be computed in polynomial time, e.g., when the metric space is a tree metric~\cite{lolakapuri2019} or the social network is a grid~\cite[Chapter 9]{ishizuka2022tfnp}.

Recently, median-based and ordinal aggregation rules have received increasing attention. Updates using the median were motivated and studied in the French-DeGroot model in~\cite{Mei2022,MeiTAC24}. Apart from convergence aspects, these works also provide structural and complexity results for consensus properties in equilibrium. More recently, some works have started to analyze median aggregation also in the Friedkin-Johnsen model~\cite{ZhangTAC25,ZhangSCIS2025}. Closer to our work is~\cite{LiCCDC22}, who simulate HKS with weighted median aggregation and explore consensus and opinion fragmentation properties. In a different direction, the computational and convergence aspects of median aggregation have also been studied in a variant that combines local and global aggregation~\cite{BerenbrinkAIJ2026}.

\section{Model and Notation}
\label{sec:model}
\paragraph{Hegselmann-Krause Systems.}
We study a generalized Hegselmann-Krause system (\agg-HKS) $(n,\opns{1},\agg)$, defined as follows.
Given a set of $n$ agents $[n]:=\lbrace 1,2,...,n \rbrace$, each agent $i \in [n]$ has an \emph{initial opinion} $\opn{i}{1} \in \R$. %
Agents update their opinions in discrete time steps $t \in \N$.
We denote the opinion of agent $i$ at the beginning of step $t$ by $\opn{i}{t}$. We denote by $\opns{t}$ the \emph{configuration} or \emph{state} at the beginning of round $t$, i.e., the vector of opinions of all $n$ agents.
Given a constant \emph{confidence bound} $\eps \in \R_{+}$ the \emph{influencing neighborhood} of agent $i$ is defined as
\[
\nh{i}{t} = \{j \in [n] : \abs{\opn{i}{t} - \opn{j}{t}} \leq \eps\}.
\]
Using appropriate scaling, we w.l.o.g.\ assume $\eps = 1$. In particular, this is a standard assumption in the literature on HKS with average aggregation. In each step $t$, one or more agents update their opinions based on the opinions of their influencing neighborhood, i.e., the multiset of \emph{neighbor opinions} $\nopns{i}{t} = \lmulti \opn{j}{t} \mid j \in \nh{i}{t} \rmulti$.
The update is performed via the aggregation function $\agg$ by choosing $\opn{i}{t+1} = \agg (\nopns{i}{t})$. We call an agent $i$ \emph{stable} at time step $t$ if $\opn{i}{t} = \agg (\nopns{i}{t})$ (and \emph{unstable} otherwise). If every agent is stable at time $t$, the HKS has converged, and we call the state $\opns{t}$ an \emph{equilibrium}. We are interested in the  \emph{convergence time} of the system, i.e., the number of steps that are required until the HKS reaches an equilibrium.

In a game-theoretic sense, the aggregation function expresses the preferences of each agent given the current state of the system. An update step for agent $i$ represents a best response of $i$ to the state $\opns{t}$ at time $t$. We address the convergence time in two variants of the HKS: the \emph{synchronous} and the \emph{asynchronous} variant. For the synchronous variant, we study simultaneous best-response dynamics -- in each time step, all unstable agents perform an update simultaneously. For the asynchronous one, we consider sequential best-response dynamics -- exactly one unstable agent $i \in [n]$ is chosen for the update. We say that the updating agent is \emph{activated} at time $t$.

We further distinguish based on the choice of the activated agent between \emph{worst-case}, \emph{best-case}, and \emph{average-case} (or \emph{expected}) convergence time. In the worst/best case, in each step an unstable agent is activated in order to make the convergence time as large/small as possible, respectively. In the average case, the activated agent is chosen uniformly at random from the set of unstable agents.

\paragraph{Aggregation Functions.}
We concentrate on ordinal aggregation functions $\agg$. A function $\agg$ is \emph{ordinal} if there exists a function $f:\N \to \N$ with $1 \leq f(m) \leq m$ for all $m \in \N$ such that for every multiset $X$ of cardinality $m$, we have $\agg(X) = x_{(f(m))}$, where $x_{(1)} \geq x_{(2)} \geq \dots \geq x_{(m)}$ is the ordered list of entries of $X$. We call $f$ the \emph{rank selector}. We call an ordinal function $\agg$ \emph{monotone} if its rank selector is monotone increasing, i.e., $f(m+1)-f(m) \ge 0$ for $m \ge 1$. A rank selector $f$ is \emph{$k$-Lipschitz} if $0 \leq f(m+1) - f(m) \leq k$ for every $m \ge 1$.
Below, we extend the definition of $k$-Lipschitz to the aggregation function $\agg$: $\agg$ is $k$-Lipschitz if and only if the corresponding rank selector $f$ is $k$-Lipschitz. %
Note that, by definition, we assume a $k$-Lipschitz function is ordinal and monotone. We say that a function $\agg$ is \emph{strictly} $k$-Lipschitz if it is $k$-Lipschitz but not $(k-1)$-Lipschitz.

For example, \minagg-aggregation has rank selector $f(m) = m$, \maxagg-aggregation has $f(m) = 1$, \largeagg{k} has $f(m) = \min\{k,m\}$, and the lower and upper median have rank selectors $f(m) = \lceil (m+1)/2 \rceil$ and $f(m) = \lfloor (m+1)/2 \rfloor$. All of these examples are $1$-Lipschitz functions.

For a HKS with ordinal $\agg$, we call a subset $A \subseteq [n]$ of agents \emph{isolated} at time $t$ if $\nh{i}{t} \subseteq A$ for all $i \in A$.
Since agents adopt opinions only from their influencing neighborhood, we have that $\opn{i}{t+1} \in \nopns{i}{t}$. Therefore, any isolated set of agents at time $t$ remains isolated at any time $t' \ge t$.

\paragraph{\largeagg{k} and \smallagg{k}.}
Our results for \largeagg{k} also apply to \smallagg{k} aggregation functions, which select the $\min\{k,|\nh{i}{t}|\}$-smallest opinion from $\nopns{i}{t}$. The definition of \largeagg{k} functions does not immediately include \smallagg{k} ones -- given the formal framework above, a \smallagg{k} aggregation has rank selector $f(m) = \max\{m-k+1,1\}$.

It is easy to verify that, by symmetry, all our proof ideas for \largeagg{k} aggregation carry over to \smallagg{k} aggregation if we consider opinion values sorted in non-decreasing order (which becomes the standard definition of order statistics). More generally, any function that is $1$-Lipschitz with the decreasing ordering (as defined above) is also $1$-Lipschitz for the natural variant with the increasing ordering. However, functions that are $k$-Lipschitz for $k \ge 2$ with one ordering are not monotone with the reverse ordering.%

\paragraph{Ordered Opinion Graphs.}
Ordinal aggregation is discrete: every agent always adopts an opinion that already appears in its influencing neighborhood. Hence, no new opinion values are created during the dynamics. In particular, the set of opinion values present in the system can only shrink over time; once an opinion value is no longer held by any agent, it disappears from the system and cannot reappear in later time steps.
This constitutes a qualitative difference from the standard averaging-based HKS. Averaging may create opinion values that were not present in the initial state, and hence equilibrium clusters need not be located at initial opinions. In contrast, ordinal aggregation guarantees that every opinion occurring throughout the dynamics, and in particular every opinion occurring in an equilibrium, belongs to the initial support. Long-run outcomes therefore arise by changing the multiplicities of initially present opinions and possibly eliminating some of them, rather than by creating new compromise opinions.

Moreover, ordinal aggregation depends only on the relative order of the opinions in a neighborhood and on their multiplicities, but not on the exact numerical values. Consequently, the evolution of the system is fully determined by the combinatorial structure induced by the initial opinion vector $\opns{1}$. For every time $t$, let $\dcto{t}:=|\{\opn{i}{t}\mid i\in[n]\}|$ denote the number of distinct opinions, so $\dcto{1}=\ell$. To express the relevant combinatorial structure, we define the undirected \emph{ordered opinion graph} $G = (V, E)$ as follows. Let $\disopn{1} < \disopn{2} < \dots <\disopn{\ell}$ be the distinct opinions in $\opns{1}$. Then $V = [\ell]$, where $u<v$ corresponds to $\disopn{u} < \disopn{v}$, and $E$ is given by undirected edges $\{u,v\} \in E$ for each pair of vertices $u,v \in V$ for which $|\disopn{u} -\disopn{v}| \le 1$. We denote the closed neighborhood of a vertex $v$ by $\onh{v} = \{v\} \cup \{ u \in V \mid \{u,v\} \in E\}$.

Observe that the ordered opinion graph admits a natural geometric representation as a unit interval graph. Each vertex $v$ can be cast as a unit interval centered at the opinion value $\disopn{v}$. Then two vertices are adjacent if and only if their corresponding unit intervals intersect.

Each vertex $v$ is initially equipped with a multiplicity $n_1(v)$, which represents the number of agents currently holding that opinion. Since updates only move agents to existing opinion values, the dynamics can equivalently be viewed as a token process on $G$. Each agent $i$ is a token. If the agent has opinion $\disopn{v}$ at time $t$, its token is located at vertex $v$. It is easy to see that the ordinal aggregation of $i$ can be performed based on the information in $G$ along with the vector of multiplicities $\mathbf{n}_t$. In particular, the influence neighborhood is generated by all agents located at neighboring opinions $u$ of $v$. The neighbor opinions $\nopns{i}{t}$ contain each opinion $\disopn{u}$ in the neighborhood of $v$ exactly $n_t(u)$ times. As such, the choice $\agg(\nopns{i}{t})$ can be executed by counting agents in increasing order of their opinions.

This equivalent view of the ordinal HKS as a token process on an ordered unit-interval graph is useful for some of our proofs below.

\section{Asynchronous Dynamics}\label{sec:async}
\subsection{Best-Case and Average-Case Convergence}

Let us first consider asynchronous HKS with ordinal aggregation. We start by showing that for every initial configuration $\opns{1}$, there is a sequence of updates that leads the HKS into an equilibrium. Hence, in a game-theoretic sense, the HKS is \emph{weakly-acyclic}~\cite{Young1993,Marden2009}. Moreover, the best-case convergence time is always polynomial.

\begin{theorem}\label[theorem]{thm:acyclic}
    Every asynchronous HKS with an ordinal aggregation function is weakly-acyclic.
    The best-case convergence time is at most $(n-1)(\ell-1)$.
\end{theorem}

We first establish the following technical lemma.

\begin{lemma}\label[lemma]{lem:equaltarget}
    Let $A_v\coloneqq \{i \in [n] \mid \opn{i}{t} = v\}$ be the set of all agents holding opinion $v$ at time $t$.
    Then:
    \begin{compactenum}
        \item $\agg(\nopns{i}{t}) = \agg(\nopns{j}{t})$ for all $i,j \in A_v$, and
        \item if $i \in A_v$ is activated at time $t$, then $\agg(\nopns{j}{t+1}) = \agg(\nopns{j}{t})$ for every $j\in A_v \setminus \{i\}$.
    \end{compactenum}
\end{lemma}

\begin{proof}
The first statement is a simple observation -- since $i$ and $j$ are located at the same opinion, they have the same influencing neighborhood $\nh{i}{t} = \nh{j}{t}$, which implies $\agg(\nopns{i}{t})=\agg(\nopns{j}{t})$.

For the second statement, let $i, j\in A_v$.
Since $\nh{i}{t} = \nh{j}{t}$, denote by $X\coloneqq \nopns{i}{t} = \nopns{j}{t}$ their common multiset of neighbor opinions and let $m=\abs{X}$.
For the ordinal function \agg, consider the rank selector $f$.
\space Let $w = \agg(X)= x_{(f(m))}$ be the $f(m)$-th largest element of $X$.
When agent $i$ is activated at time $t$, they change their opinion from $v$ to $w$.
Since $w\in X$, we have $|w-v|\le 1$, and hence $i$ remains in the influencing neighborhood of every $j\in A_v\setminus\{i\}$.
As no other opinion changes, $\nh{j}{t+1}=\nh{j}{t}$, and $\nopns{j}{t+1}$ is obtained from $X$ by replacing one occurrence of $v$ by $w$.
Replacing $v$ by $w$ can only weakly decrease the number of elements strictly larger than $w$, while it can only weakly increase the number of elements at least as large as $w$.
Hence, since initially
$|\{x\in X: x>w\}|<f(m)\le|\{x\in X: x \ge w\}|$,
these inequalities continue to hold after the replacement.
Therefore, $w$ remains the $f(m)$-th largest element so $\agg( \nopns{j}{t+1}) = \agg(\nopns{j}{t})$ as claimed.
\end{proof}

\begin{proof}[Proof of \cref{thm:acyclic}]
    Suppose the state $ \opns{t}$ is not an equilibrium. There exists at least one agent $i$ with $v = \opn{i}{t} \neq \agg(\nopns{i}{t}) = w$. By the previous lemma, all agents from $A_v = \{ j \mid \opn{j}{t} = v\}$ are unstable.

    By \cref{lem:equaltarget}, if we activate all agents in $A_v$ sequentially, they all switch to the same opinion $w$. We call this sequence of activations a \emph{phase}. In particular, at the end of the phase, every agent in $A_v$ has left opinion $v$, and the number of distinct opinions in the state decreases by one. Since the aggregation function is ordinal, the set of opinions adopted by the agents cannot expand over time.

    Initially, there are $\ell \le n$ opinions. Hence, after at most $\ell-1$ phases, the HKS must have converged. Each phase requires activating exactly $|A_v| \le n-1$ agents, who all deviate to the same target opinion. As such, if two agents have the same opinion at time $t$, they have the same opinion at the end of the current phase, and at the end of every subsequent phase. In every phase, two previously distinct opinion classes are merging. This implies an upper bound of $(n-1)(\ell-1)$ steps and proves the theorem.
\end{proof}

While the best-case convergence time is always at most quadratic in $n$, it does not directly imply a good worst-case or average-case convergence time. Indeed, it is easy to construct an ordinal HKS in which the worst-case convergence time is infinite, i.e., there exist cyclic sequences of deviations. Moreover, we show that the average-case convergence time can be exponentially large. In particular, the following proposition is based on a class of instances in which the convergence time with uniform random activation is exponential in $n$ with overwhelming probability. Interestingly, the instances have only $\ell = 3$ opinions.

\begin{proposition}\label[proposition]{prop:expavg}
For odd $n \ge 3$, there is an ordinal HKS with $n$ agents and the following property. Suppose $T$ is the convergence time of the HKS with uniform random activation of unstable agents. There exists a constant $c>0$ such that
\[
\Pr[T \ge 2^{c n}] \ge 1-2^{-\Omega(n)}.
\]
The average-case convergence time is $\mathbb{E}[T]=2^{\Omega(n)}$.
\end{proposition}

\begin{proof}
    Consider an ordinal HKS with $n$ agents and a rank selector $f$ defined by $f(s)=1$ for $1 \le s \le n-1$ and $f(s)=n$ for $s \ge n$. Let $m = (n-1)/2$. The HKS has the following initial opinions: $m$ agents with opinion $0$, $m$ agents with opinion $1$, and a single agent with opinion $2$. As long as there are agents with both opinion $0$ and $1$, every agent with opinion $0$ wants to adopt opinion $1$ and vice versa. The single agent with opinion $2$ remains stable unless all other agents adopt opinion $1$. In the latter case, the agent makes one final move to adopt 1. Thus, the HKS reaches an equilibrium (within at most one additional step) if and only if all agents of one opinion in $\{0,1\}$ adopted the other opinion and one of the two opinions vanishes. Let $T$ be the first time at which either opinion $0$ or $1$ disappears. $T$ is a random variable.

    Let $A_t$ and $B_t$ be random variables describing the number of agents holding opinions $0$ and $1$ at any time $t$, respectively. Let $X_t = |m - A_t| = |m - B_t|$. Then $X_t \in \{0,1,\dots,m\}$ and $X_0=0$. At each step $t < T$, one of the $n-1$ unstable agents with opinion $0$ or $1$ is selected uniformly at random. If a $0$-agent is selected, it moves to $1$. Then if $A_t \le m$, this increases $X_t$ by $1$, otherwise it decreases $X_t$ by 1. Selecting a $1$-agent decreases $X_t$ by $1$ if $A_t \le m$ and increases it by 1 otherwise. Then, for any $1 \le k \le m-1$
    \begin{align*}
        \Pr[X_{t+1}=k+1 \mid X_t=k] &= \frac{m-k}{n-1} = \frac12 - \frac{k}{n-1}, \\
        \Pr[X_{t+1}=k-1 \mid X_t=k] &= \frac{m+k}{n-1} = \frac12 + \frac{k}{n-1}.
    \end{align*}

    Moreover, if $X_t=0$ then $\Pr[X_{t+1}=1]=1$. An equilibrium is reached only if $X_t = m$. Thus $(X_t)$ evolves as a random walk on the integers, and the expected convergence time is at least the hitting time of $m$. The claim of the lower bound follows from \cref{lem:chain} below.
\end{proof}

\begin{lemma}\label[lemma]{lem:chain}
Let $(X_t)_{t \ge 0}$ be the Markov process on $\{0,1,\dots,m\}$ with $m = (n-1)/2$ and transition probabilities
\[
\Pr[X_{t+1} = k+1 \mid X_t = k] = \tfrac12 - \tfrac{k}{n-1}, \quad
\Pr[X_{t+1} = k-1 \mid X_t = k] = \tfrac12 + \tfrac{k}{n-1},
\]
for $1 \le k \le m-1$, $\Pr[X_{t+1} =1 \mid X_t = 0] = 1$, and $X_0 = 0$. Then there exists a constant $c > 0$ such that
\[
\Pr\big( T \leq 2^{c n} \big) = 2^{-\Omega(n)},
\]
where $T := \inf\{ t \geq 0 \mid X_t = m \}$.
\end{lemma}

\begin{proof}
We apply the negative drift theorem of Oliveto and Witt~\cite{DBLP:journals/algorithmica/OlivetoW11,DBLP:journals/corr/OlivetoW12}.
Define an interval $[a,b]$ by
\[
a = \left\lceil \frac{n-1}{6} \right\rceil, \qquad b = \left\lfloor \frac{n-1}{3} \right\rfloor.
\]
We first compute the drift. For any $k \in \{1,\dots,m-1\}$,
\[
\mathbb{E}[X_{t+1} - X_t \mid X_t = k]
= \Big(\tfrac12 - \tfrac{k}{n-1}\Big) - \Big(\tfrac12 + \tfrac{k}{n-1}\Big)
= -\frac{2k}{n-1}.
\]
Hence for all $k \in [a,b]$,
\[
\mathbb{E}[X_{t+1} - X_t \mid X_t = k] \le -\frac{2a}{n-1} \le -\varepsilon
\]
for some constant $\varepsilon > 0$.
Next, we verify the jump condition. Since $|X_{t+1} - X_t| = 1$ always,
\[
\Pr\big(|X_{t+1} - X_t| \ge j \mid X_t = k\big) = 0 \quad \text{for all } j \ge 2,
\]
and is at most $1$ for $j=0,1$. Thus the exponential tail condition required by the theorem holds for any function $r(\ell) \ge 1$, where $\ell = b-a = \Theta(n)$. Choose $r(l) =1$.

We can therefore apply the negative drift theorem, which yields that there exists a constant $c_1 > 0$ such that for
\[
T_b = \inf\{ t \ge 0 : X_0 \le a \mid X_t \ge b \},
\]
it holds
\[
\Pr\big( T_b \le 2^{c_1 \ell /r(\ell)} \big) = 2^{-\Omega(\ell/r(\ell))}.
\]
Since $\ell = \Theta(n)$ and $r(\ell)=1$, this becomes
\[
\Pr\big( T_b \le 2^{c n} \big) = 2^{-\Omega(n)}
\]
for some constant $c > 0$.
Finally, observe that in order to reach $m$, the process must first reach $b$. Therefore,
\[
\Pr(T \le 2^{c n}) \le \Pr(T_b \le 2^{c n}) = 2^{-\Omega(n)}. \qedhere
\]
\end{proof}

\subsection{Potential Functions and Worst-Case Convergence}

\Cref{thm:acyclic} establishes only weak acyclicity: from every state there exists a best-response sequence leading to an equilibrium. \Cref{prop:expavg} shows that in general best-response sequences are not guaranteed to converge. In this section, we study a number of important classes for which we obtain convergence in the worst case -- every asynchronous best-response sequence converges. We first consider \medianagg-HKS and construct a potential function that is inherently based on metric distances. In the subsequent section, we consider \largeagg{k}-HKS and an argument that exploits the interval structure of the ordered opinion graph. Finally, for the general class of \textsf{$1$-Lipschitz}-HKS we provide a potential function that combines normalized distances and linear interval structure. We prove that the existence of potential functions is not guaranteed on higher levels of the Lipschitz hierarchy. In particular, \emph{there is} a strictly $2$-Lipschitz aggregator that allows us to construct a HKS with a cyclic sequence of best responses. Moreover, \emph{every} strictly $k$-Lipschitz aggregator with $k \ge 3$ allows the construction of a HKS with a cyclic sequence of best responses.

\paragraph{\medianagg-HKS.}
For \medianagg-HKS, our convergence proof uses a potential function that strictly decreases in every update step. It initially depends on $\delta$, the smallest distance among the initial numerical opinion values.
We then combine it with a structural normalization for ordered opinion graphs to obtain a polynomial bound on the convergence time that depends solely on $n$ and $\ell$.

\begin{theorem}
    \label{thm:median}
    Every asynchronous \medianagg-HKS has a worst-case convergence time of $\O(n^2\ell)$.
\end{theorem}

We start with a discussion of the potential function.
\begin{lemma}\label[lemma]{lem:medianPotential}
    Every asynchronous \medianagg-HKS has a distance-based potential function. It yields a worst-case convergence time of $\O(n^2/\delta)$, where $\delta := \min\{ |\opn{i}{1} - \opn{j}{1}| : \opn{i}{1} \neq \opn{j}{1}\}$.
\end{lemma}
\begin{proof}
For a state $\opns{t}$ we define the \emph{individual potential} of agent $i$ as
$$
\ipot{i}{t}=\sum_{j\in[n]}\min\bigl(\{|\opn{i}{t}-\opn{j}{t}|,1\}\bigr),
$$
and the potential as $\ipot{}{t} = \sum_{i\in[n]}\ipot{i}{t}$. Each ordered pair contributes at most~$1$, so $\ipot{}{1}\le n(n-1)$. We recall a classical property of any median: For every multiset $X\subset\R$, every median element $y = \medianagg(X)$ minimizes the sum of the 1-norm distances to the elements in $X$, i.e., $y\in\arg\min_{z\in\R}\sum_{x\in X}|x-z|$.

Suppose in step $t$ we activate agent $i$. Let $\mov{i}=|\opn{i}{t+1}-\opn{i}{t}|$ denote the absolute change of the opinion of $i$ in this step. In terms of the potential, only contributions involving both $\opn{i}{t}$ and $\opn{j}{t}$ can change, for some $j \in [n]\setminus\{i\}$. Moreover, the change is symmetric in the contributions of $\ipot{i}{t}$ and $\ipot{j}{t}$. Overall, it is straightforward to verify that
$$
\ipot{}{t}-\ipot{}{t+1}
=2\sum_{j\in[n]\setminus\{i\}}\Bigl(\min\{|\opn{i}{t}-\opn{j}{t}|,1\}-\min\{|\opn{i}{t+1}-\opn{j}{t}|,1\}\Bigr).
$$
For $j\notin \nh{i}{t}$, we have $\min\{|\opn{i}{t}-\opn{j}{t}|,1\}=1$. The term is non-negative; therefore, removing it preserves a valid lower bound of the potential change.
For $j\in \nh{i}{t}$ we have
$\min\{|\opn{i}{t}-\opn{j}{t}|,1\}=|\opn{i}{t}-\opn{j}{t}|$ and always $\min\{|\opn{i}{t+1}-\opn{j}{t}|,1\}\le |\opn{i}{t+1}-\opn{j}{t}|$, thus
$$
\ipot{}{t}-\ipot{}{t+1}\ge 2\sum_{j\in \nh{i}{t}\setminus\{i\}}\Bigl(|\opn{i}{t}-\opn{j}{t}|-|\opn{i}{t+1}-\opn{j}{t}|\Bigr).
$$
Any median minimizes the 1-norm of distances, hence
$$
\sum_{j\in  \nh{i}{t}}|\opn{i}{t+1}-\opn{j}{t}|\le \sum_{j\in  \nh{i}{t}}|\opn{i}{t}-\opn{j}{t}|.
$$
Rearranging gives $\sum_{j\in  \nh{i}{t}}\bigl(|\opn{i}{t}-\opn{j}{t}|-|\opn{i}{t+1}-\opn{j}{t}|\bigr)\ge 0$. For $j \in \nh{i}{t}$ with $j=i$, we have $|\opn{i}{t}-\opn{i}{t}|-|\opn{i}{t+1}-\opn{i}{t}|=-|\opn{i}{t+1}-\opn{i}{t}|$, implying
$$
\sum_{j\in \nh{i}{t}\setminus \{i\}}\Bigl(|\opn{i}{t}-\opn{j}{t}|-|\opn{i}{t+1}-\opn{j}{t}|\Bigr)\ge |\opn{i}{t+1}-\opn{i}{t}|=\mov{i}.
$$
Combining the two equations yields $\ipot{}{t}-\ipot{}{t+1}\ge 2 \mov{i}$. This also covers the case of an even-sized neighborhood in which an agent moves from one median opinion to the other. Although both median opinions minimize the total $1$-norm distance over $\nh{i}{t}$, the self-contribution of agent $i$ equals $-\mov{i}$, and hence removing this contribution still yields the strict decrease above.

Since the \medianagg\ aggregation is ordinal, every opinion always coincides with some opinion in the initial state. Thus, whenever an unstable agent moves, we have $\mov{i} \ge \delta$, and the potential decreases by at least $\ipot{}{t}-\ipot{}{t+1}\ge 2\delta$. Since $\ipot{}{1}\le n(n-1)$, the number of steps is in $\O(n^2/\delta)$.
\end{proof}

In order to obtain a bound on the convergence time in terms of $n$ and $\ell$, we prove a lower bound on the minimum step size $\delta$. We exploit the representation of the HKS as an ordered opinion graph. The following lemma shows that every ordered opinion graph induced at time $t$ has a corresponding realization in which all distinct opinion values are separated by at least $1/\dcto{t}\geq 1/\ell$.
Thus, without loss of generality, we may assume such a minimum gap when reasoning about step sizes.

\begin{lemma}
\label[lemma]{lemma:minGap}
Let $G=(V,E)$ be the ordered opinion graph induced by a state $\opns{t}$, and let $n_t(v)$ denote the multiplicity of vertex $v\in V$. Then there exists a realization $\topns{t}\in\R^n$ of the opinion vector that induces the same ordered opinion graph $G$ and the same multiplicities $n_t(v)$ such that $\topn{i}{t} = a_i/\dcto{t}$ for $a_i \in \N$. Consequently, the minimum distance between any two distinct opinion values satisfies
$$
\delta = \min\{|\topn{i}{t} - \topn{j}{t}| : \topn{i}{t} \neq \topn{j}{t} \} \ge \frac{1}{\dcto{t}} \geq \frac{1}{\ell}.
$$
\end{lemma}

\begin{proof}

Let $G=(V,E)$ be the ordered opinion graph induced by the state $\opns{t}$. By definition, $|V|=\dcto{t}\leq\ell$. As argued above, $G$ is a unit interval graph: each vertex $v\in V$ corresponds to a unit interval centered at the opinion value $\disopn{v}$, and two vertices are adjacent exactly when the corresponding intervals intersect. %

Corneil et al.~\cite{CORNEIL199599} show that every unit interval graph with $\dcto{t}$ vertices admits a representation by intervals $[q_v,q_v+1]$ whose endpoints are rational numbers with denominator $\dcto{t}$.
Moreover, the intervals can be ordered so that $q_1 < q_2 < \dots < q_{\dcto{t}}$.

We construct a new realization of the opinion profile by assigning to vertex $v$ the opinion value $\disopn{v} = q_v$ and placing $n_t(v)$ agents at this value. Since the intersection pattern of the intervals is unchanged, this state induces the same ordered opinion graph $G$. Furthermore, all values $q_v$ are integer multiples of $1/\dcto{t}$, and hence any two distinct opinions $\disopn{u}, \disopn{v}$ differ by at least $1/\dcto{t}\geq 1/\ell$. This proves the lemma.
\end{proof}

\begin{proof}[Proof of \cref{thm:median}]
Combining \cref{lem:medianPotential,lemma:minGap}, it suffices to focus on instances with $\delta\ge 1/\ell$. Substituting this bound yields $\O(n^{2}/\delta) = \O(n^2 \ell)$ and proves the theorem.
\end{proof}

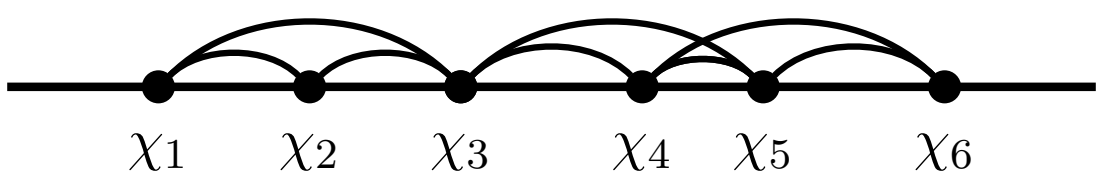
\begin{figure}[ht!]
\centering
\begin{tikzpicture}[
    x=2cm,
    y=1cm,
    >=Latex,
    opinion/.style={circle,fill=black,inner sep=2.2pt},
    depedge/.style={very thick,black,shorten <=3pt,shorten >=3pt},
]

\def\xA{0.0}
\def\xB{0.5}
\def\xC{1.0}
\def\xD{1.6}
\def\xE{2.0}
\def\xF{2.6}

\draw[ultra thick] (-0.5,0) -- (3.1,0);

\foreach \x/\lab in {
    \xA/\disopn{1},
    \xB/\disopn{2},
    \xC/\disopn{3},
    \xD/\disopn{4},
    \xE/\disopn{5},
    \xF/\disopn{6},
}{
    \node[opinion] at (\x,0) {};
    \node[below=5pt] at (\x,0) {$\lab$};
}

\draw[depedge] (\xA,0) to[bend left=45] (\xB,0); %
\draw[depedge] (\xB,0) to[bend left=45] (\xC,0); %
\draw[depedge] (\xA,0) to[bend left=45] (\xC,0); %
\draw[depedge] (\xC,0) to[bend left=45] (\xD,0); %
\draw[depedge] (\xD,0) to[bend left=45] (\xE,0); %
\draw[depedge] (\xC,0) to[bend left=45] (\xE,0); %
\draw[depedge] (\xD,0) to[bend left=45] (\xE,0); %
\draw[depedge] (\xD,0) to[bend left=45] (\xF,0); %
\draw[depedge] (\xE,0) to[bend left=45] (\xF,0); %

\end{tikzpicture}
\caption{Structure of the instance in Example~\ref{ex:median}. Location of opinions; arrows indicate edges in the ordered opinion graph \label{fig:median}}
\end{figure}

\begin{example} \label{ex:median} \rm
Consider a HKS with the $\ell = 6$ opinions $0.0, 0.5, 1.0, 1.6, 2.0, 2.6$, see Figure~\ref{fig:median}. Suppose in step $t$ there is a single agent at every opinion. Observe that there is a minimum distance of $0.4 > 1/\ell$ between distinct opinions. The potential is $\ipot{}{t} = 25.2$. Now suppose agents adopt an upper median. Then agent 4 wants to move from $1.6$ to $2.0$. This creates an improvement in the distance between agents (4,5) and (4,6) of 0.4 each. The distance between agents (3,4) increases by 0.4. Since all distances are counted twice, the potential decreases by $4\cdot 0.4 - 2\cdot 0.4 = 0.8$, which is exactly twice the moving distance. Now suppose that agent 6 moves from $2.6$ to $2.0$. Then the distance between agents (4,6) and (5,6) decreases by 0.6 each, so the potential decreases by $4\cdot 0.6 = 2.4$, which is more than twice the moving distance. \hfill $\blacksquare$
\end{example}

\paragraph{\largeagg{k}-HKS.}
For \medianagg aggregation, we can bound the worst-case convergence time by combining numerical properties of the median with a suitable graph-theoretic representation of the instance. For \largeagg{k} aggregation, we again resort to the graph-theoretic representation and exploit a different monotonicity structure using a notion of load for cuts between neighboring opinions.

\begin{theorem} \label{thm:k-large-potential}
    Every asynchronous \largeagg{k}-HKS has a worst-case convergence time of $\O(n \ell^2)$.
\end{theorem}

We prove the theorem by constructing an integer-valued potential on the ordered opinion graph.
We may assume $1 \le k \le n$, since every $k > n$ induces the same aggregation function as $k=n$. Consider the ordered opinion graph $G$ induced by the initial state $\opns{1}$, with $V = [\ell]$ according to the opinion order. Recall that $\onh{v}$ is the closed neighborhood of $v$. If $v+1 \not\in \onh{v}$, then no edge crosses between $\{1,\ldots,v\}$ and $\{v+1,\dots, \ell\}$. Hence, the ordered opinion graph decomposes at these cuts. Since agents only adopt opinions from their influencing neighborhoods, distinct components evolve independently. We may therefore assume that $G$ is connected, which implies $v+1 \in \onh{v}$ for every $v \in [\ell-1]$.

For every vertex $v \in [\ell]$, let $r(v) = \max\{u: u \in \onh{v}\}$ be its rightmost neighbor. The function $r$ is non-decreasing. Moreover, by connectivity $r(v) \ge v+1$ for all $v \in [\ell-1]$. We call the edge $\{v,v+1\}$ the \emph{cut} of $v$. Recall that $n_t(v)$ denotes the number of agents located at vertex $v$ at the beginning of step $t$. Consider the load of cut $v$ as the number of neighboring agents with strictly higher opinions, i.e.,
\begin{equation}
    \label{eq:load}
    \load{v}{t} = \sum_{\substack{u \in \onh{v}\\u>v}} n_t(u) = \sum_{u=v+1}^{r(v)}n_t(u).
\end{equation}
An update from vertex $u$ to vertex $w$ crosses cut $v$ from left to right if $u\leq v<w$; crossing from right to left is defined similarly.

Intuitively, when $\load{v}{t} \geq k$, any agent that crosses the cut of $v$ must cross from left to right -- the $k$-th largest neighborhood opinion lies strictly right of the cut, so no agent with an opinion larger than $v$ would want to deviate to a lower opinion. Such a deviation increases $\load{v}{t}$ by 1. Similarly, $\load{v}{t} \le k-1$ means any agent can only cross from right to left, which decreases $\load{v}{t}$ by 1. Thus, to measure progress towards equilibrium, we count the minimum distance of the load from the pair $\{k-1,k\}$. In particular, $d(j) = j-k$ for $j \ge k$ and $d(j) = k-1-j$ for $j \le k-1$.

\begin{lemma} \label[lemma]{lem:cut-local}
    Consider a cut $v\in [\ell-1]$ at time $t$.
    \begin{compactenum}
        \item If the activated agent crosses cut $v$, then $d(\load{v}{t+1}) = d(\load{v}{t}) +1$.
        \item If the activated agent does not cross cut $v$, then $d(\load{v}{t+1})$ can decrease only if the agent crosses cut $r(v)$, and in this case it decreases by at most $1$.
    \end{compactenum}
\end{lemma}

\begin{proof}
    Let agent $i$ be activated in step $t$, and let $u,w\in[\ell]$ be its current and target opinions, respectively. Thus, after the update, the token of $i$ has moved from $u$ to $w$.

    Suppose first that $i$ crosses cut $v$ from left to right, i.e., $u\le v < w$. Then $|\nh{i}{t}|\ge k$. Indeed, if $|\nh{i}{t}|<k$, \largeagg{k} selects the smallest opinion in $\nopns{i}{t}$, which cannot be strictly larger than the current opinion of $i$. Hence, the target at vertex $w$ is the $k$-th largest opinion in $\nopns{i}{t}$, and at least $k$ agents in $\nh{i}{t}$ are located at vertices greater than or equal to $w$. Every such vertex belongs to $\onh{u}$ and is therefore at most $r(u) \le r(v)$. Consequently, all these agents are counted by $\load{v}{t}$, and thus $\load{v}{t} \ge k$. Moreover, the activated agent enters the interval of vertices $\{v+1,\dots, r(v)\}$, so $\load{v}{t+1}= \load{v}{t} +1$. Since $\load{v}{t} \ge k$, it follows that $d(\load{v}{t+1}) = d(\load{v}{t}) +1$.

    Suppose now that $i$ crosses the cut $v$ from right to left, i.e., $w \le v < u$. Since $u$ and $w$ are adjacent, $u\in \onh{v}$, and hence $u \le r(v)$. Moreover, every vertex in $\{v+1,\dots,r(v)\}$ belongs to $\onh{u}$. Indeed, $u$ is adjacent to $v$ because $w\leq v<u$ and $u$ is adjacent to $w$. Hence, every vertex from $v+1$ through $u$ is adjacent to $u$. Every vertex from $u$ through $r(v)$ is also adjacent to $u$, since its distance to $u$ is no larger than its distance to $v$, which is at most $1$.
    Therefore, all agents counted by $\load{v}{t}$ belong to $\nh{i}{t}$ and hold opinions strictly larger than the target opinion at vertex $w$. If $\load{v}{t} \ge k$, then the $k$-th largest opinion in $\nopns{i}{t}$ would lie strictly to the right of cut $v$, contradicting $w \le v$. Hence, $\load{v}{t} \le k-1$. The activated agent leaves the interval $\{v+1, \dots, r(v)\}$, so $\load{v}{t+1} = \load{v}{t}-1$. Since the load was at most $k-1$, this decrease raises its distance from $\{k-1,k\}$ by one. Thus, $d(\load{v}{t+1}) = d(\load{v}{t}) +1$, proving the first statement.

    Finally, suppose the update does not cross cut $v$. Since $\load{v}{t}$ counts exactly the agents at vertices $v+1, \dots, r(v)$, it can change only if the update crosses the other boundary, cut $r(v)$. One update changes $\load{v}{t}$ by one and hence decreases $d(\load{v}{t})$ by at most one, proving the second statement.
\end{proof}

The lemma already allows to define a strict lexicographic potential, so asynchronous dynamics cannot cycle. Consider a step $t$ and the rightmost cut $v$ crossed in this step. For every cut $u>v$, neither cut $u$ nor cut $r(u)>u$ is crossed, so $d(\load{u}{t})$ remains unchanged. At cut $v$, the value $d(\load{v}{t})$ increases by one. Hence, the distance vector $(d(\load{\ell-1}{t}),\dots, d(\load{1}{t}))$ strictly increases lexicographically.

To obtain a polynomial bound on the convergence time, we provide a suitable aggregation of the distance vector into an integral potential function of bounded size.

\begin{proof}[Proof of \cref{thm:k-large-potential}]
    Construct a directed forest on the cuts $[\ell-1]$ by making cut $r(v)$ the parent of cut $v$ whenever $r(v) < \ell$. Since $r(v) >v$, every edge points to a  cut with a strictly larger index. If $r(v)=\ell$, then cut $v$ is a root.

    For every cut $v$, we define its weight recursively by
    \[
     a_v = 1 + \sum_{\substack{q\in [\ell-1]\\ r(q)=v}} a_q.
    \]
    Equivalently, $a_v$ is the number of cuts in the subtree rooted at $v$. Every cut in this subtree has index at most $v$, and hence $a_v \le v$.

    Consider the non-negative integer potential
    \[
    \ipot{}{t} = \sum_{v=1}^{\ell-1} a_v d(\load{v}{t}).
    \]
    When an agent moves from opinion $u$ to $w$ in step $t$, and denote the set of crossed cuts by $A_t=\{v\in[\ell-1]:\min\{u,w\}\leq v<\max\{u,w\}\}$. By \cref{lem:cut-local}, every cut in $A_t$ contributes an increase of $a_v$ to the potential. A cut $q\notin A_t$ can contribute a decrease only if $r(q)\in A_t$, and this decrease is at most $a_q$. Moreover, an update cannot cross both cut $v$ and cut $r(v)$. Hence,
    \[
        \ipot{}{t+1}-\ipot{}{t}
        \ge \sum_{v\in A_t}a_v
        -\sum_{\substack{q\in[\ell-1]\\r(q)\in A_t}}a_q
        =\sum_{v\in A_t}
        \bigg(a_v-\sum_{\substack{q\in[\ell-1]\\r(q)=v}}a_q\bigg)
        =|A_t|=|u-w|\ge 1.
    \]
    Thus, the potential strictly increases in every step.

    For every $j\in\{0,\ldots,n\}$, we have $0\leq d(j)\leq n-1$. Together with $a_v\leq v$, this gives
    \[
        0\leq\ipot{}{t}
        \leq(n-1)\sum_{v=1}^{\ell-1}a_v
        \leq(n-1)\sum_{v=1}^{\ell-1}v
        =\frac{(n-1)\ell(\ell-1)}{2}
    \]
    when $G$ is connected. Otherwise, we apply the construction individually to every connected component and sum the resulting potentials. This can only improve the upper bound due to the quadratic dependence on $\ell$.
\end{proof}

\begin{figure}[ht!]
\centering
\begin{tikzpicture}[
    x=2cm,
    y=1cm,
    >=Latex,
    opinion/.style={circle,fill=black,inner sep=2.2pt},
    depedgeA/.style={-stealth,very thick,blue!75!black,shorten <=3pt,shorten >=3pt},
    depedgeB/.style={-stealth,very thick,teal!70!black,shorten <=3pt,shorten >=3pt}
]

\def\xA{0.0}
\def\xB{0.5}
\def\xC{1.0}
\def\xD{1.6}
\def\xE{1.9}
\def\xF{2.6}

\draw[ultra thick] (-0.5,0) -- (3.1,0);

\foreach \x/\lab in {
    \xA/\disopn{1},
    \xB/\disopn{2},
    \xC/\disopn{3},
    \xD/\disopn{4},
    \xE/\disopn{5},
    \xF/\disopn{6},
}{
    \node[opinion] at (\x,0) {};
    \node[below=5pt] at (\x,0) {$\lab$};
}

\draw[depedgeA] (\xA,0) to[bend left=45] (\xC,0); %
\draw[depedgeA] (\xB,0) to[bend left=45] (\xC,0); %
\draw[depedgeA] (\xC,0) to[bend left=45] (\xE,0); %
\draw[depedgeA] (\xD,0) to[bend left=45] (\xF,0); %
\draw[depedgeA] (\xE,0) to[bend left=45] (\xF,0); %
\end{tikzpicture}

\vspace{1cm}

\begin{tikzpicture}[
    >=Latex,
    cut/.style={circle,draw,thick,minimum size=8mm},
    terminal/.style={circle,draw,thick,minimum size=8mm,fill=gray!15},
    edge/.style={->,thick,shorten <=2pt,shorten >=2pt},
    weight/.style={font=\small}
]

\node[cut]      (r1) at (0,0)     {$1$};
\node[cut]      (r2) at (2,0)   {$2$};
\node[cut]      (r3) at (1,1.5) {$3$};
\node[cut]      (r4) at (3,3) {$4$};
\node[cut]      (r5) at (1,3) {$5$};

\draw[edge] (r1) -- (r3);
\draw[edge] (r2) -- (r3);
\draw[edge] (r3) -- (r5);

\node[weight, left=3pt of r1] {$a_1=1$};
\node[weight, right=3pt of r2] {$a_2=1$};
\node[weight, left=3pt of r3] {$a_3=3$};
\node[weight, right=3pt of r4] {$a_4=1$};
\node[weight, left=3pt of r5] {$a_5=4$};

\end{tikzpicture}
\caption{Structure of the instance in Example~\ref{ex:kLarge}. Top: Location of opinions, arrows indicate $(v,r(v))$ for all $v$ with $r(v) > v$. Bottom: Directed forest of cuts. Node labels indicate the cut. \label{fig:kLarge}}
\end{figure}
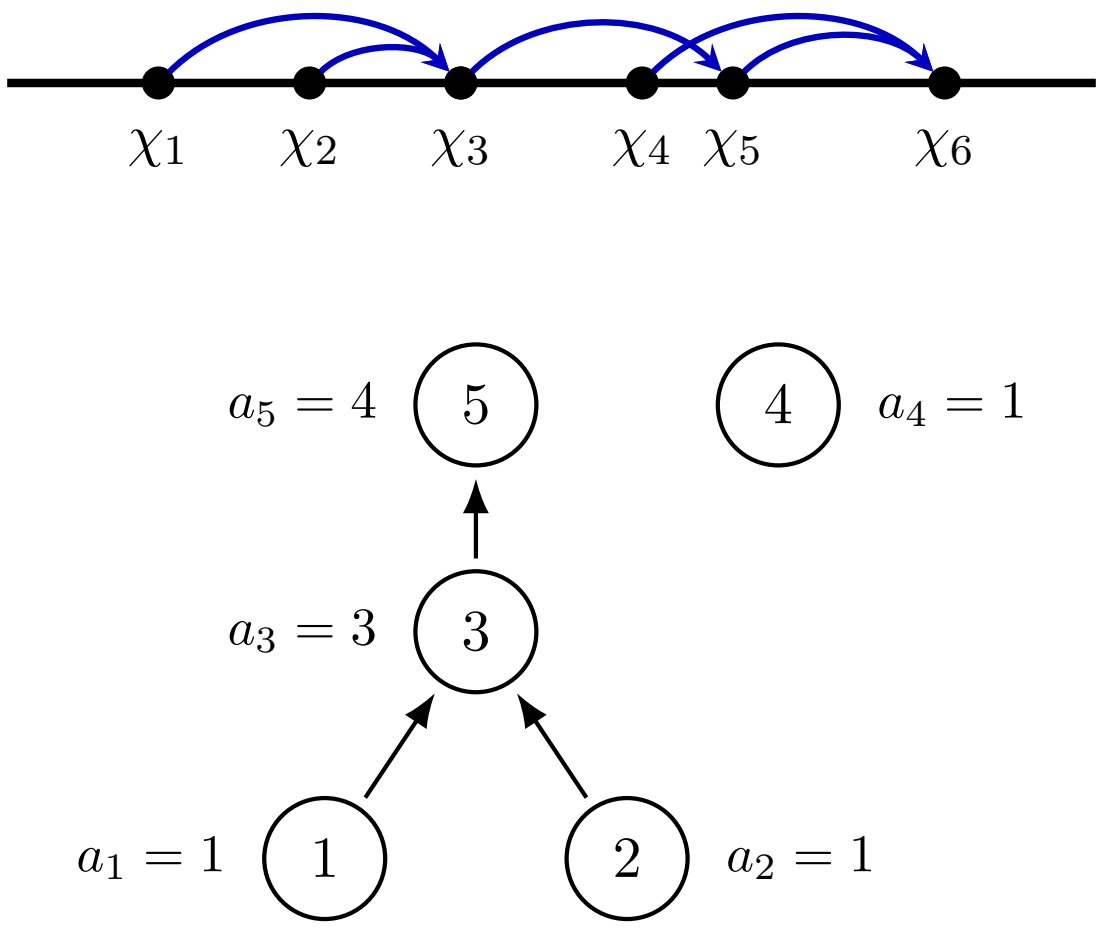

\begin{example} \label{ex:kLarge} \rm
Reconsider the instance from Example~\ref{ex:median} with 6 initial opinions $0.0, 0.5, 1.0, 1.6, 2.0, 2.6$; see Figure~\ref{fig:kLarge}. Suppose that for each opinion there are 2 agents holding that opinion. The loads of the cuts $1,\ldots,5$ are $\loads{t} = (4,2,4,4,2)$.

First, suppose $k = 2$. Then the distance values are $d(\loads{t}) = (2,0,2,2,0)$. As such, the value of the potential is $\ipot{}{t} = \sum_{v=1}^5 a_v d(\load{v}{t}) = 10$. Now suppose an agent from $u = \disopn{3}$ moves to $w = \disopn{5}$. The new values for $\loads{t+1} = (3,1,5,5,2)$ and $d(\loads{t+1}) = (1,0,3,3,0)$, so $\ipot{}{t+1} = 13$, and $\ipot{}{t+1} - \ipot{}{t} = 13-10 = 3 > 2 = 5-3$.

Second, consider $k = 3$. Then the distance values are $d(\loads{t}) = (1,0,1,1,0)$. As such, the value of the potential is $\ipot{}{t} = \sum_{v \in [6]} a_v d(\load{v}{t}) = 5$. Now suppose an agent from $u = \disopn{3}$ moves to $w = \disopn{4}$. The new values for $\loads{t+1} = (3,1,5,4,2)$ and $d(\loads{t+1}) = (0,1,2,1,0)$, so $\ipot{}{t+1} = 8$, and $\ipot{}{t+1} - \ipot{}{t} = 8-5 > 4-3$.  \hfill $\blacksquare$
\end{example}

For \maxagg-aggregation, it is easy to observe that only $\O(n \ell)$ steps will be made in the worst case, since each agent can increase their opinion in the ordered opinion graph at most $\ell-1$ times. A symmetric argument applies for \minagg-aggregation.

\begin{observation}
    \label{obs:minimax}
   Every asynchronous HKS with \minagg- or \maxagg-aggregation has a worst-case convergence time of $\O(n \ell)$.
\end{observation}

\paragraph{\textsf{$1$-Lipschitz}-HKS.}

Both \medianagg and \largeagg{k} aggregation functions are $1$-Lipschitz. Notably, this common structural property is sufficient to guarantee convergence of asynchronous dynamics in the worst case, even in polynomial time. The proof adapts and generalizes the cut-load approach used in the proof of Theorem~\ref{thm:k-large-potential} above, but the general rank selector requires a more flexible way of assigning values to the loads. As in the proof for \largeagg{k}-HKS, we construct an integer-valued potential from loads assigned to cuts. Agent updates cross consecutive cuts and change these loads by one. For \largeagg{k}, the fixed threshold $k$ allowed to evaluate every cut load by a single function $d(i)$ expressing the distance to the set $\{k-1,k\}$ and then assign tree-based weights to cuts. For a general $1$-Lipschitz selector, the profitable direction depends on both the loads to the left and right of the crossed cut. We therefore carefully assign a marginal distance value to each additional unit of load. Once these marginal distance values have been constructed, the potential again simply sums integer values of cut distances.

\begin{theorem} \label{thm:lipschitz-potential}
    Every asynchronous HKS with a $1$-Lipschitz ordinal aggregation function has a worst-case convergence time of $\O(n^2 \ell^2)$.
\end{theorem}

\begin{proof}
Rather than just numbering opinions from 1 to $\ell$ and using cuts between consecutive opinions, we compose a more careful numerical representation that preserves the geometric structure. By \cref{lemma:minGap}, we replace the initial opinions by equivalent rational coordinate values whose denominators are $\ell$. W.l.o.g.\ we again concentrate on a single connected component of the ordered opinion graph $G$. We multiply all opinions by $\ell$ to make them integral. Accordingly, this also scales the confidence bound by $\ell$. The instance then has integral opinion values from $\{0,\ldots,\ell(\ell-1)\}$ and an integral confidence bound of $\ell$. To highlight the connections to the earlier proof for \largeagg{k} aggregation, we shift the instance by another additive $\ell$, i.e., opinions are integers $V \subset \{\ell,\ldots,\ell^2\}$.

We now consider a cut $\{v, v+1\}$ for every single integer $v = 0,\ldots,\ell^2-1$, even if the value does not represent an opinion in $V$.
Extending the earlier multiplicity notation, let $n_t(u)$ denote the number of agents at integer opinion $u$, with $n_t(u)=0$ if no agent holds that opinion. %
The load, in direct analogy to \eqref{eq:load}, is given by
\[
    \load{v}{t} = \sum_{u = v+1}^{v+\ell} n_t(u).
\]

Suppose an agent best responds to the right from $u$ to $w > u$. For every crossed integer cut $v\in\{u,\ldots,w-1\}$, the agent enters the load of cut $v$ and leaves the load of cut $v-\ell$. A left move performs the reverse transfer. The two loads count agents in the adjacent intervals $(v-\ell,v]$ and $(v,v+\ell]$, each of exactly one confidence length. This gives a one-to-one correspondence between lost and gained load. In particular, a simple sum over all loads remains constant.

Let $a = \load{v-\ell}{t}-1$ and $b = \load{v}{t}$ be the numbers of agents other than the mover at cuts $v-\ell$ and $v$, respectively. The two intervals are disjoint, so $a+b+1\leq n$. We seek non-negative integer marginal rewards $\mu_{r,s}$ for $r = 0,\ldots,\ell-2$ and $s = 0,\ldots,n$ such that increasing the load of cut $v$ from $b$ to $b+1$ has value $\mu_{\lfloor v/\ell\rfloor,b}$. Intuitively, we define uniform marginal rewards for blocks of $\ell$ consecutive cuts. Since the paired cuts $v-\ell$ and $v$ belong to different, consecutive blocks, their indices differ by $\lfloor v/\ell\rfloor - \lfloor (v-\ell)/\ell\rfloor = 1$. With $r = \lfloor v/\ell\rfloor$, when the agent moves right, we remove $\mu_{r,a}$ and add $\mu_{r+1,b}$, while a move to the left does the reverse. Each of these should yield an improvement if the moving agent does a best response. In the following two lemmas we establish the existence of suitable integer rewards $\mu_{r,s}$. In particular, we show that
\begin{eqnarray}
    \text{best response to the right} \ \Longrightarrow\ & b\geq f(a+b+1)
    &\ \Longrightarrow\
    \mu_{r+1,b}\geq\mu_{r,a}+1
    \label{eq:lipschitz-right-marginal}\\
    \text{best response to the left} \ \Longrightarrow\ & b\leq f(a+b+1)-2
    &\ \Longrightarrow\
    \mu_{r,a}\geq\mu_{r+1,b}+1,
    \label{eq:lipschitz-left-marginal}
\end{eqnarray}
where the second set of implications will hold for all $0 \le r \le \ell-2$. We start by establishing the first implications, which are numerical consequences of best responses in terms of the rank selector.

\begin{lemma}\label[lemma]{lem:lipschitz-cut-loads}
Consider a non-trivial update from $u$ to $w$ and an integer cut $v$ crossed by the update. Let $a$ and $b$ be the loads of cuts $v-\ell$ and $v$, excluding the moving agent. If $w>u$, then $b\geq f(a+b+1)$. If $w<u$, then $b\leq f(a+b+1)-2$.
\end{lemma}

\begin{proof}
Let $m$ be the size of the moving agent's neighborhood. First suppose that $w>u$, so $u\leq v<w$. Since $w$ is the $f(m)$-th largest neighborhood opinion, at least $f(m)$ neighbors have opinion at least $w$. Note that $v<w\leq u+\ell\leq v+\ell$, so each such agent is therefore counted by $\load{v}{t}$. The mover is not among these agents, and hence $b\geq f(m)$. Conversely, every agent counted by $\load{v-\ell}{t}$ has opinion in $(v-\ell,v]$. Since $u\leq v<w\leq u+\ell$, this agent is at a neighboring opinion of $u$ strictly smaller than $w$. Including the mover, there are $a+1$ such agents, so $a+1\leq m-f(m)$.
Overall, if $a+b+1\leq m$, monotonicity gives $f(a+b+1)\leq f(m)\leq b$. If $a+b+1>m$, the $1$-Lipschitz property gives $f(a+b+1)\leq f(m)+a+b+1-m\leq b$. Thus $b\geq f(a+b+1)$.

Now suppose that $w<u$, so $w\leq v<u$. At least $m-f(m)+1$ neighbors have opinion at most $w$. Their opinions lie in $(v-\ell,v]$ and are therefore counted by $\load{v-\ell}{t}$. The mover is not among them, and hence $a\geq m-f(m)+1$. Every other agent counted by $\load{v}{t}$ has opinion in $(v,v+\ell]$. Since $w\leq v<u$ and $u-w\leq\ell$, it is a neighbor with opinion strictly larger than $w$. Together with the mover, there are $b+1$ such agents. Since at most $f(m)-1$ neighbors have an opinion strictly larger than the selected opinion $w$, we obtain $b+1\leq f(m)-1$.
Together, if $a+b+1\geq m$, monotonicity gives $f(a+b+1)\geq f(m)\geq b+2$. If $a+b+1<m$, applying the $1$-Lipschitz property backwards gives $f(a+b+1)\geq f(m)-(m-a-b-1)=f(m)-m+a+b+1\geq b+2$. Thus $b\leq f(a+b+1)-2$.

The cases $a+b+1>m$ for a rightward move and $a+b+1<m$ for a leftward move are the key point: here $1$-Lipschitzness transfers information about the neighborhood size $m$ to the load value $a+b+1$.
\end{proof}

Our second lemma turns the numerical conditions of best responses into a suitable set of marginal rewards that implies a strict increase whenever a best response is performed.

\begin{lemma}\label[lemma]{lem:lipschitz-marginals}
There are non-negative integers $\mu_{r,s}<\ell n$, for $0 \leq r \leq \ell-1$ and $0\leq s<n$, satisfying \eqref{eq:lipschitz-right-marginal} and \eqref{eq:lipschitz-left-marginal} for all $a,b\geq0$ with $a+b+1\leq n$.
\end{lemma}

\begin{proof}
For every integer $b\in\{0,\ldots,n-1\}$, define
\[
    T(b) =
    \max\Bigl(\{-1\}\cup
    \{a\in\{0,\ldots,n-1\}:b\geq f(a+b+1)\}\Bigr)
    +\frac{b+1}{n+1}.
\]
The integer part records the largest opposing load $a$ for which the rightward comparison is required at load $b$; the fractional part only breaks ties between consecutive values of $b$. The maximum is non-decreasing in $b$. Indeed, if $b\geq f(a+b+1)$, then $1$-Lipschitzness gives $f(a+b+2)\leq f(a+b+1)+1\leq b+1$, so every $a$ admitted for $b$ is also admitted for $b+1$. The fractional term therefore gives $T(0)<T(1)<\cdots<T(n-1)$. Extend $T$ to a strictly increasing function on $\mathbb{R}$ by piecewise-linear interpolation.

The definition directly encodes the desired comparisons. If $b\geq f(a+b+1)$, then $a<T(b)$. If $b\leq f(a+b+1)-2$, then monotonicity gives $f(a'+b+1)>b$ for every $a'\geq a$. Thus no such $a'$ occurs in the maximum defining $T(b)$, and $T(b)<a$.

We assign the pair $(r,s)$ a real-valued score $T^{(r)}(s)$ by repeated application, where $T^{(0)}(s) = s$ and $T^{(r+1)}(s) = T(T^{(r)}(s))$. Recall that $T$ is strictly increasing. Thus, $a < T(b)$ and applying $T$ on both sides $r$ times implies $T^{(r)}(a)<T^{(r+1)}(b)$. Similarly, $T(b)<a$ implies the reverse strict inequality. Now sort all distinct scores of all $\ell n$ pairs $(r,s)$ from smallest to largest. Let $\mu_{r,s}$ be the position of the score of $(r,s)$ in this sorted sequence. These integer ranks lie between $0$ and $\ell n-1$. Every strict inequality of scores yields a difference of at least one between the corresponding ranks.

Thus, if $b\geq f(a+b+1)$, then $a<T(b)$, so $T^{(r)}(a) < T^{(r+1)}(b)$ and $\mu_{r,a} < \mu_{r+1,b}$ for all $0 \le r \le \ell-1$. Similarly, if $b\leq f(a+b+1)-2$, then $a > T(b)$, so $T^{(r)}(a) > T^{(r+1)}(b)$ and $\mu_{r,a} > \mu_{r+1,b}$ for all $0 \le r \le \ell-1$. This proves the conditions in \eqref{eq:lipschitz-right-marginal} and \eqref{eq:lipschitz-left-marginal}.
\end{proof}

Using the marginal rewards from \cref{lem:lipschitz-marginals}, we measure progress with the following potential function\footnote{Note that the structure of the potential suggests interpreting the HKS as an atomic congestion game~\cite{Rosenthal73} with cuts as resources and marginal rewards as delays for each cut in the corresponding block. For this game, $\ipot{}{t}$ would be an exact potential function~\cite{MondererS96} However, progress in this congestion game does not necessarily imply progress in the HKS for every possible agent move, since actual improvements in the HKS satisfy further conditions in~\Cref{lem:lipschitz-cut-loads}.}
\[
    \ipot{}{t} =
    \sum_{v=0}^{\ell^2-1}
    \sum_{s=0}^{\load{v}{t}-1}
    \mu_{\lfloor v/\ell\rfloor,s},
\]
where the inner sum is empty when $\load{v}{t}=0$. Increasing a cut load from $s$ to $s+1$ adds $\mu_{\lfloor v/\ell\rfloor,s}$, while decreasing it removes the same marginal reward.

Consider a best response of an agent from $u$ to $w \neq u$ and a crossed cut $v$. The change in $\Phi$ is
\begin{align*}
\ipot{}{t+1} - \ipot{}{t} \; = \; \sum_{v=u}^{w-1} \mu_{\lfloor(v-\ell)/\ell\rfloor+1,b}-\mu_{\lfloor(v-\ell)/\ell\rfloor,a} \; \geq \; |u-w| & \qquad \text{ for best response to the right} \\
\ipot{}{t+1} - \ipot{}{t} \; = \; \sum_{v=w}^{u-1} \mu_{\lfloor(v-\ell)/\ell\rfloor,a}-\mu_{\lfloor(v-\ell)/\ell\rfloor+1,b} \; \geq \; |u-w| & \qquad \text{ for best response to the left}
\end{align*}
by \cref{lem:lipschitz-cut-loads} and \eqref{eq:lipschitz-left-marginal}. Thus, every update increases the integer potential by at least $|u-w|\geq1$.

Finally, every agent is counted by exactly the $\ell$ cut loads immediately to the left of its opinion. Hence $\sum_{v=0}^{\ell^2-1}\load{v}{t}=n\ell$. Since every marginal reward is non-negative and smaller than $\ell n$,
\[
    0\leq\ipot{}{t}
    < \ell n
    \sum_{v=0}^{\ell^2-1}\load{v}{t}
    =\ell^2 n^2.
\]
Thus, every sequence of improvement steps has length at most $\ell^2 n^2$. Due to the quadratic dependence on $n$ and $\ell$, the asymptotic upper bound continues to apply when we consider each connected component of $G$ separately and sum the corresponding worst-case convergence times in each component.
\end{proof}

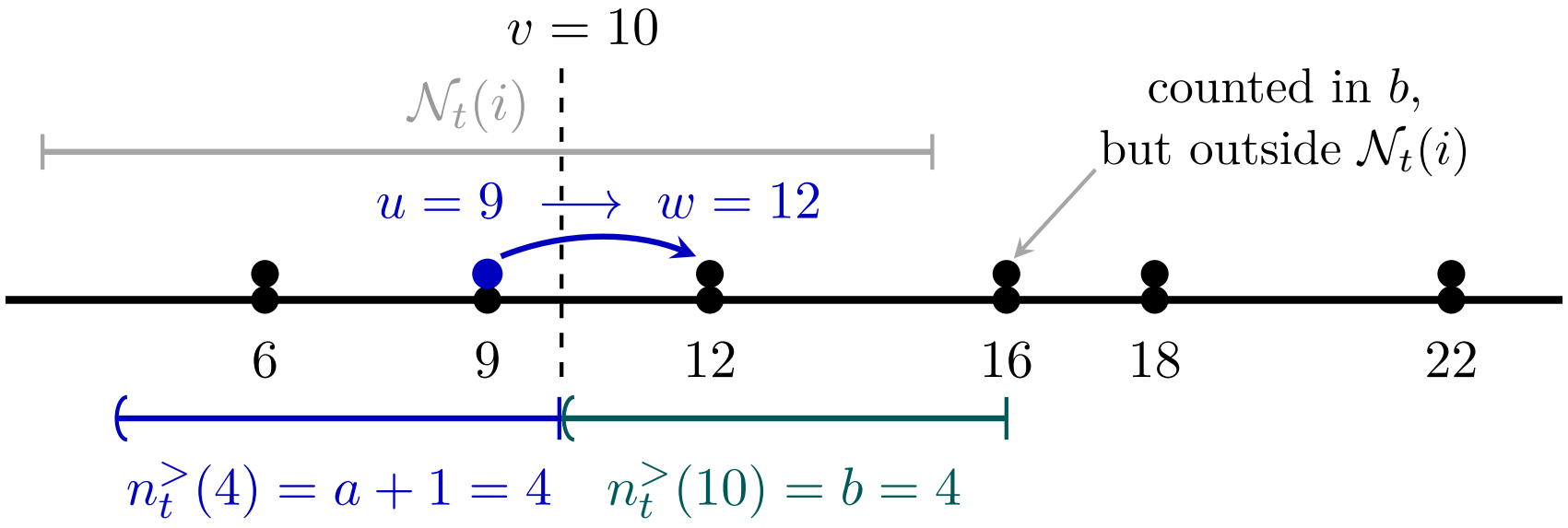
\begin{figure}[ht!]
\centering
\begin{tikzpicture}[
    x=0.38cm,
    y=0.85cm,
    xscale=1.4,
    >=Latex,
    opinion/.style={circle,fill=black,inner sep=2pt},
    mover/.style={circle,fill=blue!75!black,inner sep=2.2pt},
    move/.style={-stealth,very thick,blue!75!black,
                 shorten <=3pt,shorten >=3pt},
]

\draw[ultra thick] (2.5,0) -- (23.5,0);

\foreach \x/\lab in {
    6/6,
    9/9,
    12/12,
    16/16,
    18/18,
    22/22
}{
    \node[opinion] at (\x,0) {};
    \node[opinion] at (\x,0.22) {};
    \node[below=5pt] at (\x,0) {$\lab$};
}

\node[mover] at (9,0.22) {};

\draw[move] (9,0.32) to[bend left=30]
    node[above] {$u=9\ \longrightarrow\ w=12$} (12,0.32);

\draw[very thick,gray!70] (3,1.25) -- (15,1.25);
\draw[thick,gray!70] (3,1.10) -- (3,1.40);
\draw[thick,gray!70] (15,1.10) -- (15,1.40);
\node[above] at (8.75,1.25) {\textcolor{gray!80}{$\nh{i}{t}$}};

\draw[dashed,thick] (10,-0.65) -- (10,2.0);
\node[above right] at (9,2.0) {$v=10$};

\def\yint{-1}
\def\yrad{0.18}
\def\xrad{0.14}

\draw[very thick,blue!75!black]
    (4,\yint) -- (10,\yint);

\draw[thick,blue!75!black]
    (4.14,\yint+\yrad)
    arc[start angle=90,end angle=270,
        x radius=\xrad,y radius=\yrad];

\draw[thick,blue!75!black]
    (10-0.03,\yint-\yrad) -- (10-0.03,\yint+\yrad);

\node[below=5pt,blue!75!black] at (7,\yint)
    {$\load{4}{t}=a+1=4$};

\draw[very thick,teal!70!black]
    (10,\yint) -- (16,\yint);

\draw[thick,teal!70!black]
    (10.14+0.03,\yint+\yrad)
    arc[start angle=90,end angle=270,
        x radius=\xrad,y radius=\yrad];

\draw[thick,teal!70!black]
    (16,\yint-\yrad) -- (16,\yint+\yrad);

\node[below=5pt,teal!70!black] at (13,\yint)
    {$\load{10}{t}=b=4$};

\node[align=center,anchor=west] at (17,1.5)
    {\small counted in $b$,\\[-1pt]\small but outside $\nh{i}{t}$};
\draw[-stealth,thick,gray!70] (17.2,1.1) -- (16.10,0.35);

\end{tikzpicture}
\caption{Equivalent integer realization in Example~\ref{ex:lipschitz}.
The agent at $u=9$ best responds to $w=12$. For the representative
crossed cut $v=10$, the paired loads count the adjacent confidence-length
intervals $(v-\ell,v]$ and $(v,v+\ell]$. The agents at opinion $16$ are
counted in the latter load although they are outside the moving agent's
neighborhood, yielding $a+b+1=8>m=6$.
\label{fig:lipschitz}}
\end{figure}

\begin{example}\label{ex:lipschitz}\rm
Reconsider the instance from Example~\ref{ex:median} with 6 initial
opinions $0.0,0.5,1.0,1.6,2.0,2.6$, and suppose that two agents hold each
opinion. Consider the $1$-Lipschitz rank selector
$f(m)=\lceil\frac{m}{3}\rceil$.
Using the equivalent integer realization $6,9,12,16,18,22$ with confidence bound $\ell=6$, an
agent at $u=9$ has a neighborhood of size $m=6$ and best responds to $w=12$,
since $f(6)=2$.

The update crosses the cuts $v=9,10,11$. The corresponding values $(a,b)$
from \cref{lem:lipschitz-cut-loads} are $(3,2), (3,4), (3,4),$ respectively; see Figure~\ref{fig:lipschitz}. In particular, at $v=10,11$,
we have $a+b+1=8>m=6$, since the cut loads also count the two agents at
opinion $16$. Here the $1$-Lipschitz property yields $f(8)\le f(6)+(8-6)=4=b$.

For this selector, a simple way to instantiate the marginal rewards explicitly is 
giving $\mu_{r,s}$ the rank $2^r(s+1)-1$ among all marginal types
$(r,s)$. Indeed, the right- and leftward conditions reduce to
$a\le 2b-1$ and $a\ge 2b+3$, respectively, so these ranks satisfy
\eqref{eq:lipschitz-right-marginal}--\eqref{eq:lipschitz-left-marginal}. 
We use this simple ranking\footnote{Using an explicit application of \Cref{lem:lipschitz-marginals}, we give different pairs $(r,s)$ the same rank if their numerical value stemming from the application of $T$ coincides. This would yield smaller ranks of $\mu_{0,3}=3, \mu_{1,2}=5,\mu_{1,4}=9$.} for exposition: $\mu_{0,3}=4, \mu_{1,2}=8,\mu_{1,4}=16$. Consequently, the three crossed cuts contribute
\newcommand{\blacksquarehere}{\tag*{$\blacksquare$}}
\begin{equation*}
    \ipot{}{t+1}-\ipot{}{t}
    =(8-4)+(16-4)+(16-4)
    =28
    >|w-u|=3. \blacksquarehere
\end{equation*}
\end{example}

\paragraph{Lipschitz Hierarchy.}
Intuitively, for $1$-Lipschitz aggregation functions, best responses do not change very drastically when small changes happen in their neighborhood. We next observe that when best responses vary more significantly, convergence is not guaranteed. While the best-case bound from \cref{thm:acyclic} still applies, we can construct cyclic best response sequences, i.e., the worst-case convergence time can become unbounded, and there is no strict potential function.

\begin{proposition}\label[proposition]{thm:nopotential}
    Consider any aggregation function \agg with a rank selector $f$ that satisfies $f(m) = p$ and $f(m+j) \geq p+j+2$ for some $m \geq p,j \in \N$. Then there is an asynchronous \agg-HKS with a two-cycle, i.e., there is no strict ordinal potential function in this instance.
\end{proposition}

\begin{proof}
    Consider $n = m+j$ agents. Initially, there are $m-p$ agents at opinions $0$, $p$ agents at opinion 1, and $j$ agents at opinion $2$.  Suppose agent $1$ is at opinion 0. At time $t=1$, it has an influence neighborhood of size $m$. It will move to opinion $1$. If at time $t=2$ agent $1$ is activated again, the neighborhood has size $m+j$, and it will move back to $0$ -- there are $j$ agents with opinion $2$ and $p+1$ agents with opinion $1$, so the $p+j+2$-largest one is $0$. Thus, the state returns to the initial one after two steps. This shows that no potential function exists and asynchronous dynamics can cycle.
\end{proof}

The conditions of the proposition are trivially satisfied by any $k$-Lipschitz function with $k \ge 3$, since there exist $m > p\in \N$ such that $f(m) = p $ and $f(m+1)\geq p+3$. In the class of strictly $2$-Lipschitz functions, we next outline one example fulfilling the conditions of \cref{thm:nopotential}. However, not all $2$-Lipschitz functions yield cyclic behavior under these initial conditions. It remains open whether for each strictly $2$-Lipschitz function there exist initial conditions that give rise to cyclic behavior.

Consider a HKS with 6 agents and a 2-Lipschitz aggregator $f(i)$ that is $1,1,1,1,3,5$ for $i=1,\ldots,6$, respectively. Then the proposition applies using $m=4$, $p=1$, $j=2$, i.e., $f(m) = p$ and $f(m+j) \ge p+j+2$ hold since $f(4)=1$ and $f(6)=5$.

\begin{corollary}
    There exists a HKS and an initial state such that asynchronous dynamics can cycle and no strict potential exists
    \begin{compactenum}
        \item for at least one strictly $2$-Lipschitz aggregation function, and
        \item for every strictly $k$-Lipschitz aggregation function with $k \geq 3$.
    \end{compactenum}
\end{corollary}

Note that together with \cref{thm:nopotential} this provides a tight characterization of acyclicity with respect to the hierarchy of $k$-Lipschitz functions.

\section{Synchronous Hegselmann--Krause Systems}\label{sec:sync}

We now turn to synchronous HKS with ordinal aggregation. Our first result shows that there is a large set of natural ordinal aggregation functions that are guaranteed to reach an equilibrium in $\Theta(n)$ steps. These functions preserve the relative order of agents' opinions in each update step. We call an aggregation function \emph{order-preserving} if it satisfies
\begin{align*}
    \agg(X) \leq \agg\bigl(X\setminus \{\min(X)\}\bigr) \qquad \text{ and } \qquad \agg(X) \geq \agg\bigl(X \setminus \{\max(X) \}\bigr),
\end{align*}
for all multisets $X$ with $|X| \geq 2$.

First, we show that the relative order of opinions is preserved in each update step for any synchronous HKS with an ordinal order-preserving aggregation function.
\begin{lemma}
\label[lemma]{lemma:dop}
    For any synchronous HKS with an ordinal order-preserving aggregation function,
$\opn{i}{t} \leq \opn{j}{t} \, \implies\, \opn{i}{t+1} \leq \opn{j}{t+1}$ for all agents $i,j \in [n]$ and all $t \geq 1$.
\end{lemma}

\begin{proof}
Assume $i,j \in [n]$ with $\opn{i}{t} \leq \opn{j}{t}$ and set $X_0=\nopns{i}{t}$.
When shifting the confidence interval from $\opn{i}{t}$ to $\opn{j}{t}$, the opinions can leave the interval only through its left endpoint and enter only through its right endpoint. Consequently, the opinions in $\nh{i}{t} \setminus \nh{j}{t}$ can be removed successively as the current minima, while the opinions in $\nh{j}{t} \setminus \nh{i}{t}$ can be added successively as current maxima. Thus, there exists a finite sequence of non-empty multisets.
$X_0, X_1, \dots, X_m= \nopns{j}{t}$ where each step is either
\begin{compactenum}%
\item[(i)] removing a current minimum: $X_{s+1}=X_s\setminus\{\min(X_s)\}$, or
\item[(ii)] adding an element not smaller than the current maximum: $X_{s+1}=X_s\cup\{y\}$ with $y\geq \max(X_s)$.
\end{compactenum}
By the order-preserving property, in case~(i) $\agg(X_s) \leq \agg(X_{s+1})$.
In case~(ii), the added element~$y$ is a maximum of $X_{s+1}$, and hence $\agg(X_{s+1}) \geq \agg\bigl(X_{s+1}\setminus\{\max(X_{s+1})\}\bigr)=\agg(X_s)$.
Consequently,
\[
\opn{i}{t+1}
= \agg(X_0)
\leq \agg(X_m)
= \opn{j}{t+1},
\]
as claimed.
\end{proof}

We exploit this structured behavior of opinions to show that at each time step, prior to convergence, an opinion vanishes.
\begin{lemma}
\label[lemma]{cl:distinctOpinions}
In every step of a synchronous HKS with an ordinal order-preserving aggregation function, either an equilibrium is reached, or the number of distinct opinions strictly decreases.
\end{lemma}
\begin{proof}
Any pair of agents sharing the same opinion at time $t$ will do so at time $t+1$, as they have identical neighborhoods and hence receive identical updates. In each update step, a group of deviating agents can remain stationary, move to an opinion that is evacuated by all other agents, or merge with another group of agents; they never split in future time steps.

Clearly, no new opinions can appear, so $\dcto{t+1} \leq \dcto{t}$.
Assume for contradiction that the system does not converge at step $t$, that is, $ \opns{t} \neq \opns{t+1}$. By \Cref{lemma:dop}, the agents perform monotonic updates within the set of opinions at time $t$. Consequently, the number of distinct opinions must decrease by at least one, i.e., $\dcto{t+1} \leq \dcto{t} - 1$.
\end{proof}

Since the number of distinct opinions is linear in the number of agents and no new opinions can emerge, \Cref{cl:distinctOpinions} directly implies a linear bound on the synchronous convergence time.

\begin{theorem}
\label{thm:dop}
Every synchronous HKS with an ordinal order-preserving aggregation function converges in at most $n - 1$ steps.
\end{theorem}
\begin{proof}%
Initially, the number of distinct opinions satisfies $\dcto{1}=\ell\leq n$.
\Cref{cl:distinctOpinions} guarantees that in each update step, the number of distinct opinions decreases by at least one.
If, at any point, all agents share the same opinion, the HKS is in a trivial equilibrium.
Thus, after at most $n - 1$ steps, any such HKS must converge.
\end{proof}

Given that order-preserving aggregation yields linear convergence time, we explore the prevalence of this property among ordinal aggregation functions. It turns out that they represent exactly the class of $1$-Lipschitz functions. We prove the equivalence in the following theorem.

\begin{theorem}
\label{thm:orderpres_lipschitz}
    Let $\agg$ be an ordinal aggregation function. Then $\agg$ is order-preserving if and only if it is $1$-Lipschitz.
\end{theorem}
\begin{proof}
    Assume that $\agg$ is order-preserving. We first observe that every ordinal order-preserving function $\agg$ is $1$-Lipschitz.
    Consider a multiset $X=\lmulti x_1,\dots,x_m\rmulti$ whose ordered entries satisfy $ x_{(1)} > x_{(2)} > \dots > x_{(m)}$.

    First, consider removing the minimum.
    Since $\min(X)=x_{(m)}$, the ordered entries of $X\setminus\{\min(X)\}$ are $x_{(1)} > x_{(2)} > \dots > x_{(m-1)}$.
    By order-preservation, we have $ \agg(X) \leq \agg(X\setminus\{\min(X)\})$.
    Using ordinality, this gives $ x_{(f(m))} \leq x_{(f(m-1))}$.
    Since the entries of $X$ are strictly ordered, this implies $ f(m) \geq f(m-1)$.
    Hence $f(m)-f(m-1) \geq 0$, so $f$ is monotone.

    Second, consider removing the maximum.
    Since $ \max(X)=x_{(1)},$ the ordered entries of $X\setminus\{\max(X)\}$ are $ x_{(2)} > x_{(3)} > \dots > x_{(m)}$.
    Therefore, $ \agg(X\setminus\{\max(X)\}) = x_{(f(m-1)+1)}$.
    By order-preservation, we have $ \agg(X) \geq \agg(X\setminus\{\max(X)\})$.
    Using ordinality again, this yields $ x_{(f(m))} \geq x_{(f(m-1)+1)}$.
    Since the entries are strictly ordered, it follows that $ f(m) \leq f(m-1)+1$.
    Thus $f(m)-f(m-1) \leq 1$.

    Overall, we obtain $0 \leq f(m+1)-f(m) \leq 1$ for every $m \in \N$. Hence $f$ is $1$-Lipschitz.

    For the other direction, assume that $f$ is $1$-Lipschitz. We prove that $\agg$ is order-preserving.
    Let $X=\lmulti x_1 , \dots , x_m \rmulti$ be a multiset of size $m \geq 2$.
    Then \[\agg(X\setminus \{\min(X)\}) = \agg(X\setminus \{x_{(m)}\} ) = x_{(f(m-1))} \geq x_{(f(m))} = \agg(X)\] by monotonicity of $f$.
    Likewise, \[\agg(X\setminus \{\max(X)\}) = \agg(X\setminus \{x_{(1)}\} ) = x_{(f(m-1)+1)} \leq x_{(f(m))} = \agg(X)\] as $f$ is $1$-Lipschitz.
    Thus $\agg$ satisfies both order-preservation conditions. Hence, an ordinal aggregation function is order-preserving if and only if it is $1$-Lipschitz.
\end{proof}

As a consequence of \cref{thm:orderpres_lipschitz}, synchronous HKS with \minagg, \maxagg, \largeagg{k} and \medianagg aggregation all converge in at most $n-1$ steps. The linear bound is tight in general.

\begin{proposition}\label[proposition]{prop:syncOmegaN}
There are initial states for HKS with \largeagg{k} aggregation and with \medianagg aggregation on which synchronous dynamics require $\Omega(n)$ steps to converge.
\end{proposition}

\begin{proof}
    We first show the result for \largeagg{k} aggregation. Consider a chain-like instance where $\opn{i}{1} = i$ for all agents $i \in [n]$.  For $k \neq 2$, the HKS with \largeagg{k} aggregation converges in exactly $n-1$ synchronous steps. If $k=1$, only agent $n$ has no neighbor with a larger opinion and is stable. In every step $t$ all agents $i$ with $\opn{i}{t} \neq \opn{n}{1}$ adopt the opinion of their higher neighbor. At each step, exactly one additional agent adopts opinion $\opn{n}{1}$. Hence, convergence requires $n-1$ steps. If $k>2$, the behavior is symmetrical, except that the initially stable opinion is that of the lowest agent and that agents adopt the lower opinion in their neighborhood.

    \largeagg{2} aggregation needs at least $2 \cdot \lfloor n/3 \rfloor$ steps on a chain-like instance. Initially, all agents are stable except agent~$n$, who updates to $\opn{n-1}{1}$. Thus, after one step, $\opn{n-1}{2} = \opn{n}{2}$. Now, agent~$n\! -\! 2$ has two agents to their right in their neighborhood and updates towards them in step~$2$. All other agents remain stable in this step. We have $\opn{n-2}{3} = \opn{n-1}{3} = \opn{n}{3}$, and $|\opn{n-3}{3} - \opn{n-2}{3}| > 1$. Consequently, the agents in $[n-3]$ form a new chain at step~$3$. So every two steps, the group of agents forming a chain reduces by three, until convergence is reached. Hence, convergence needs at least $2 \cdot \lfloor n/3 \rfloor$ steps.

    We now proceed to show the lower bound for \medianagg aggregation. In particular, we give two examples exhibiting the required convergence time for upper and lower median aggregation. Let $n = 5c+3$ with $c \in \N$.
For the lower median $f(m) = \lceil (m+1)/2 \rceil$ the initial opinions %
are $\opn{i}{1} = \lceil 3i/5 \rceil$.
There are $3c+2$ distinct and evenly spaced opinions. There is one agent with opinion 1, then two agents with opinions 2 and 3 each, and this (1,2,2)-occupancy pattern continues towards higher opinions.
Initially, all agents are stable, but agent 1 updates to $\opn{2}{1}$.
At time step two agents $4$ and $5$ are unstable and update their opinions to $\opn{4}{3}= \opn{5}{3} = \opn{2}{2}$.
Now the lowest five agents share an opinion that is isolated, and the rest of the opinions are arranged as before for $c' = c - 1$.
So every two steps, the number of non-isolated agents reduces by five, until in the final step agent $n-2$ joins agents $n-1$ and $n$ on opinion $3c+2$. Hence, convergence needs at least $2c + 1 > n/5$ steps.

The upper median $f(m)= \lfloor (m+1)/2 \rfloor$ shows symmetrical behavior where the groups of five agents with the highest opinions isolate instead of those with the lowest opinions on inverted instances where $\opn{i}{1} = \lceil (3i-1)/5 \rceil$.
\end{proof}

Finally, we conclude with a tight characterization in the Lipschitz hierarchy. The class of $1$-Lipschitz functions always leads to convergence for synchronous HKS in finite time. We show that there are strictly $2$-Lipschitz functions for which convergence is not guaranteed.

\begin{proposition}
    There are strictly $2$-Lipschitz aggregation functions $\agg$ for which there is a synchronous $\agg$-HKS that does not converge.
\end{proposition}

\begin{proof}
    Consider the strictly $2$-Lipschitz function $\agg$ with rank selector $f(2)=1$ and $f(m)=3$ for all $m \ge 3$.
    In the initial state, there are 3 agents with $\opn{1}{1} = 0$, $\opn{2}{1} = 1$ and $\opn{3}{1}= 2$.
    Then, $\opn{1}{t+1} = \opn{2}{t}$, $\opn{2}{t+1} = \opn{1}{t}$ and $\opn{3}{t+1} = \opn{3}{t}$.
    Thus, $\opns{t}= \opns{t+2}\neq \opns{t+1}$ and the HKS does not converge.
\end{proof}

\section{Conclusion and Future Work}

We have shown that replacing the arithmetic mean by the median yields a polynomial worst-case bound for asynchronous updates. Our results identify the $1$-Lipschitz property of ordinal rank selectors as a common structural guarantee across both synchronous and asynchronous update regimes. It is equivalent to order-preserving aggregation and yields at most $n-1$ synchronous rounds, while asynchronous convergence takes $\O(n^2\ell^2)$ updates in the worst case. Median and \largeagg{k} aggregation admit sharper asynchronous bounds. This provides a theoretical justification for the median robustness in bounded-confidence models. However, ordinality alone does not exclude cycles or exponentially slow average-case convergence time for asynchronous activation.

To stimulate future analytical work, we report some insights on our preliminary simulations in the asynchronous setting. Initial opinions are sampled independently from $U[0, n/\rho]$, with $\eps = 1$; each step activates an unstable agent chosen uniformly at random. Thus $T$ counts individual updates to equilibrium, while $\rho$ controls initial spatial density. Size sweeps use 50 profiles for $n\le8000$, 30 at $n=16000$, and 20 at $n=32000$; the rank and density sweeps use $100$ profiles per setting. Rules and ranks reuse each profile, with independent activation randomness.\footnote{Simulation code and data: \url{https://github.com/lindner-jakob/ordinal-hks}.}

\begingroup
% ---------- Global size, spacing, and appearance ----------
\def\HKSPlotWidth{0.36\linewidth}
\def\HKSPlotHeight{0.24\linewidth}
\def\HKSPlotHorizontalGap{1cm}
\def\HKSPlotVerticalGap{2.8cm}
\def\HKSPlotLineWidth{0.95pt}
\def\HKSPlotMarkSize{1.8pt}
\def\HKSPlotLabelFont{\small}
\def\HKSPlotTickFont{\small}
\def\HKSPlotTitleFont{\small}
\def\HKSPlotNoteFont{\small}
\def\HKSPlotLegendFont{\small}
\def\HKSPlotLegendColumns{2}
\def\HKSPlotPlacement{ht!}

% ---------- Labels, titles, and original sweep annotations ----------
\def\HKSPlotYLabel{Mean $T/n$}
\def\HKSPlotSizeXLabel{Number of agents $n$}
\def\HKSPlotRankXLabel{$k$ in $k$-large}
\def\HKSPlotDensityXLabel{Density parameter $\rho$}
\def\HKSPlotTitleA{(a) Size sweep, $\rho=10$}
\def\HKSPlotTitleB{(b) Size sweep, $\rho=35$}
\def\HKSPlotTitleC{(c) $k$-large: rank sweep}
\def\HKSPlotTitleD{(d) Density sweep}
\def\HKSPlotNoteC{$n=10{,}000$, $x_1(i)\sim U[0,n/10]$, $\varepsilon=1$}
\def\HKSPlotNoteD{$n=10{,}000$, $x_1(i)\sim U[0,n/\rho]$, $\varepsilon=1$}
\def\HKSPlotMedianLegend{Lower median}
\def\HKSPlotFiveLegend{$5$-large}
\def\HKSPlotTwentyFiveLegend{$25$-large}
\def\HKSPlotMaxLegend{Max (1-large)}

% ---------- Common axes and curve styling ----------
\pgfplotsset{
  hksDiscussionAxis/.style={
    width=\HKSPlotWidth,height=\HKSPlotHeight,scale only axis,
    xmode=linear,ymode=linear,
    grid=both,
    major grid style={draw=gray!30},minor grid style={draw=gray!15},
    tick align=outside,tick pos=left,
    xlabel near ticks,ylabel near ticks,
    ylabel={\HKSPlotYLabel},
    label style={font=\HKSPlotLabelFont},
    tick label style={font=\HKSPlotTickFont},
    title style={font=\HKSPlotTitleFont},
    every axis plot/.append style={
      line width=\HKSPlotLineWidth,mark size=\HKSPlotMarkSize
    },
    legend image post style={line width=\HKSPlotLineWidth,mark options={solid}},
    legend style={
      draw=none,fill=none,font=\HKSPlotLegendFont,
      legend columns=\HKSPlotLegendColumns,
      /tikz/every even column/.append style={column sep=0.8em}
    }
  },
  hksDiscussionAxis/.belongs to family=/pgfplots/scale,
  hksRightAxis/.style={ytick pos=right,yticklabel pos=right},
  hksMedianPlot/.style={blue,mark=*},
  hksFivePlot/.style={orange!85!black,mark=square*},
  hksTwentyFivePlot/.style={green!50!black,mark=triangle*},
  hksMaxPlot/.style={red!75!black,mark=diamond*},
  hksRankPlot/.style={blue,mark=*}, 
% ---------- Per-panel scales, limits, and tick settings ----------
  hksPanelA/.style={
    title={\HKSPlotTitleA},xlabel={\HKSPlotSizeXLabel},
    xmode=log,log basis x=2,ymode=linear,
    xtick={250,1000,4000,16000,32000},
    xticklabels={250,1k,4k,16k,32k},
    ymin=0,ymax=24,ytick={0,5,10,15,20},scaled y ticks=false,
    legend to name=hks-discussion-legend
  },
  hksPanelA/.belongs to family=/pgfplots/scale,
  hksPanelB/.style={
    /pgfplots/hksRightAxis,
    /pgfplots/title={\HKSPlotTitleB},/pgfplots/xlabel={\HKSPlotSizeXLabel},
    /pgfplots/xmode=log,/pgfplots/log basis x=2,/pgfplots/ymode=log,
    /pgfplots/xtick={250,1000,4000,16000,32000},
    /pgfplots/xticklabels={250,1k,4k,16k,32k},
    /pgfplots/ymin=1,/pgfplots/ymax=700,
    /pgfplots/ytick={1,10,100},/pgfplots/yticklabels={1,10,100}
  },
  hksPanelB/.belongs to family=/pgfplots/scale,
  hksPanelC/.style={
    /pgfplots/clip=false,
    /pgfplots/title={\HKSPlotTitleC},/pgfplots/title style={yshift=15pt},
    /pgfplots/xlabel={\HKSPlotRankXLabel},
    /pgfplots/xmode=linear,/pgfplots/ymode=linear,/pgfplots/ymin=0,
    /pgfplots/xtick={0,10,20,30,40,50},/pgfplots/ytick={0,5,10,15,20},
    /pgfplots/scaled y ticks=false
  },
  hksPanelC/.belongs to family=/pgfplots/scale,
  hksPanelD/.style={
    /pgfplots/clip=false,
    /pgfplots/hksRightAxis,
    /pgfplots/title={\HKSPlotTitleD},/pgfplots/title style={yshift=15pt},
    /pgfplots/xlabel={\HKSPlotDensityXLabel},
    /pgfplots/xmode=linear,
    /pgfplots/ymode=log,
    %/pgfplots/ymin=0,
    /pgfplots/ymax=350,
    /pgfplots/ytick={1,10,100},
    /pgfplots/yticklabels={1,10,100},
  },
  hksPanelD/.belongs to family=/pgfplots/scale
}

% Size sweep: rho=10, rule=lower_median; x=n.
\pgfplotstableread[row sep=\\,col sep=space]{
x mean low high \\
250 1.51256 1.484318 1.538804 \\
500 1.51096 1.493036 1.528883 \\
1000 1.52284 1.509116 1.5359625 \\
2000 1.52966 1.52033825 1.53896 \\
4000 1.52195 1.51564 1.528835375 \\
8000 1.527925 1.522126688 1.5337865 \\
16000 1.52369375 1.519128958 1.528252135 \\
32000 1.525220312 1.521950977 1.528664766 \\
}\HKSScaleTenMedianData

% Size sweep: rho=10, rule=5-large; x=n.
\pgfplotstableread[row sep=\\,col sep=space]{
x mean low high \\
250 1.93912 1.874146 2.004562 \\
500 1.93316 1.895719 1.971801 \\
1000 1.96196 1.9299895 1.989921 \\
2000 2.00724 1.98295975 2.031192 \\
4000 2.003345 1.98791975 2.018495125 \\
8000 2.0055425 1.99592625 2.015831125 \\
16000 2.012708333 1.999340573 2.024214844 \\
32000 2.013992187 2.004652852 2.022959609 \\
}\HKSScaleTenFiveData

% Size sweep: rho=10, rule=25-large; x=n.
\pgfplotstableread[row sep=\\,col sep=space]{
x mean low high \\
250 6.17128 5.76593 6.605712 \\
500 6.72892 6.323941 7.151721 \\
1000 7.11616 6.8603135 7.38712 \\
2000 6.99843 6.80007575 7.20205025 \\
4000 7.2061 7.0542715 7.36975525 \\
8000 7.3371825 7.22924775 7.451274438 \\
16000 7.348941667 7.270589948 7.422567604 \\
32000 7.242473438 7.176265195 7.310132461 \\
}\HKSScaleTenTwentyFiveData

% Size sweep: rho=10, rule=max; x=n.
\pgfplotstableread[row sep=\\,col sep=space]{
x mean low high \\
250 11.11616 10.434102 11.799062 \\
500 16.07552 14.705217 17.487645 \\
1000 18.17252 17.122879 19.228322 \\
2000 20.13963 19.028457 21.40684225 \\
4000 19.919535 19.21740137 20.62507363 \\
8000 21.3255525 20.65692825 21.98087425 \\
16000 21.19631458 20.87034432 21.57680818 \\
32000 21.3270625 20.88194328 21.85308898 \\
}\HKSScaleTenMaxData

% Size sweep: rho=35, rule=lower_median; x=n.
\pgfplotstableread[row sep=\\,col sep=space]{
x mean low high \\
250 1.8956 1.851838 1.9388 \\
500 1.98608 1.949918 2.019001 \\
1000 2.01834 1.994698 2.041492 \\
2000 2.02024 2.002008 2.039911 \\
4000 2.04406 2.03535475 2.053395625 \\
8000 2.0434875 2.03312225 2.053925875 \\
16000 2.041279167 2.032534062 2.049934115 \\
32000 2.046664063 2.038324258 2.054627969 \\
}\HKSScaleThirtyFiveMedianData

% Size sweep: rho=35, rule=5-large; x=n.
\pgfplotstableread[row sep=\\,col sep=space]{
x mean low high \\
250 4.0916 4.026796 4.150176 \\
500 8.1298 8.026877 8.223364 \\
1000 16.01444 15.850739 16.1929625 \\
2000 31.87564 31.6749455 32.08231975 \\
4000 63.765045 63.36630312 64.1202685 \\
8000 125.6142125 122.8182311 127.2712804 \\
16000 244.3276292 232.7144704 254.1988065 \\
32000 490.9923 468.264388 507.6109097 \\
}\HKSScaleThirtyFiveFiveData

% Size sweep: rho=35, rule=25-large; x=n.
\pgfplotstableread[row sep=\\,col sep=space]{
x mean low high \\
250 1.8796 1.832476 1.927214 \\
500 2.02704 1.975037 2.077243 \\
1000 2.13828 2.1041345 2.173282 \\
2000 2.16351 2.13827775 2.18931075 \\
4000 2.20874 2.188899 2.22968925 \\
8000 2.2060875 2.193979062 2.218323375 \\
16000 2.2108 2.198811042 2.220859427 \\
32000 2.219729688 2.211726406 2.228442031 \\
}\HKSScaleThirtyFiveTwentyFiveData

% Size sweep: rho=35, rule=max; x=n.
\pgfplotstableread[row sep=\\,col sep=space]{
x mean low high \\
250 4.1052 4.06664 4.14328 \\
500 7.85392 7.786308 7.916938 \\
1000 15.20132 15.0988995 15.305994 \\
2000 30.11094 29.973548 30.254971 \\
4000 59.91324 59.74938262 60.08307513 \\
8000 119.3498175 119.0245341 119.6590031 \\
16000 238.0392375 237.5616493 238.4905611 \\
32000 475.2586328 474.3776845 476.1031322 \\
}\HKSScaleThirtyFiveMaxData

% Original polished rank sweep: n=10000, rho=10; x=k.
\pgfplotstableread[row sep=\\,col sep=space]{
x mean plus minus \\
1 21.271202 0.376278 0.372385 \\
2 6.934977 0.066822 0.066779 \\
3 3.713248 0.02167 0.02179 \\
4 2.525857 0.01062 0.010661 \\
5 2.009723 0.006472 0.006388 \\
8 1.613622 0.00344 0.003407 \\
10 1.611839 0.003278 0.003323 \\
12 1.686086 0.003645 0.003608 \\
15 1.994195 0.006114 0.006339 \\
20 3.446769 0.019928 0.02004 \\
25 7.303709 0.063948 0.065036 \\
30 14.459445 0.211576 0.209673 \\
40 21.428095 0.386464 0.392938 \\
50 21.208322 0.392424 0.381963 \\
}\HKSRankData

% Original polished density sweep: n=10000, rule=lower_median; x=rho.
\pgfplotstableread[row sep=\\,col sep=space]{
x mean low high \\
1 0.559004 0.558002 0.55999 \\
5 1.217387 1.215383 1.219428 \\
10 1.522758 1.520175 1.525375 \\
15 1.702671 1.698647 1.706678 \\
20 1.821919 1.817645 1.826212 \\
25 1.913345 1.908481 1.918265 \\
30 1.986732 1.981497 1.992004 \\
35 2.045363 2.039869 2.050917 \\
40 2.098127 2.092536 2.103712 \\
45 2.13503 2.128568 2.141597 \\
}\HKSDensityMedianData

% Original polished density sweep: n=10000, rule=5-large; x=rho.
\pgfplotstableread[row sep=\\,col sep=space]{
x mean low high \\
1 0.688297 0.687035 0.689575 \\
5 1.30029 1.298519 1.302165 \\
10 2.002524 1.995475 2.009545 \\
15 4.224444 4.190761 4.258645 \\
20 11.737118 11.557909 11.915886 \\
25 42.870477 41.16233 44.653169 \\
30 141.66077 134.698476 148.687325 \\
35 156.895229 154.449223 158.663469 \\
40 138.112779 137.804274 138.429555 \\
45 121.806332 121.525359 122.091606 \\
}\HKSDensityFiveData

% Original polished density sweep: n=10000, rule=25-large; x=rho.
\pgfplotstableread[row sep=\\,col sep=space]{
x mean low high \\
1 0.717328 0.715769 0.718887 \\
5 3.73792 3.717825 3.757989 \\
10 7.238748 7.171838 7.306553 \\
15 2.720705 2.707381 2.733921 \\
20 2.06156 2.055171 2.06794 \\
25 1.983638 1.978272 1.989143 \\
30 2.048858 2.042903 2.054782 \\
35 2.206175 2.197312 2.214972 \\
40 2.49539 2.484146 2.506358 \\
45 2.943825 2.927814 2.960251 \\
}\HKSDensityTwentyFiveData

% Original polished density sweep: n=10000, rule=max; x=rho.
\pgfplotstableread[row sep=\\,col sep=space]{
x mean low high \\
1 0.717246 0.715746 0.718736 \\
5 3.733802 3.712973 3.754918 \\
10 21.265503 20.947481 21.585689 \\
15 146.262814 139.222614 153.339896 \\
20 264.753481 262.956442 266.150005 \\
25 211.396146 211.002811 211.78397 \\
30 174.739747 174.441779 175.039994 \\
35 149.026644 148.809041 149.242127 \\
40 129.894978 129.71257 130.079365 \\
45 115.195759 115.024665 115.365738 \\
}\HKSDensityMaxData

\newcommand{\HKSPlotMean}[2]{%
  \addplot[#2] table[x=x,y=mean] {#1};
}

\edef\HKSStartFigure{\noexpand\begin{figure}[\HKSPlotPlacement]}
\HKSStartFigure
\centering
\begin{tikzpicture}
\begin{groupplot}[
  hksDiscussionAxis,
  group style={group size=2 by 2,
    horizontal sep=\HKSPlotHorizontalGap,vertical sep=\HKSPlotVerticalGap}
]
\nextgroupplot[hksPanelA]
\HKSPlotMean{\HKSScaleTenMedianData}{hksMedianPlot}
\addlegendentry{\HKSPlotMedianLegend}
\HKSPlotMean{\HKSScaleTenFiveData}{hksFivePlot}
\addlegendentry{\HKSPlotFiveLegend}
\HKSPlotMean{\HKSScaleTenTwentyFiveData}{hksTwentyFivePlot}
\addlegendentry{\HKSPlotTwentyFiveLegend}
\HKSPlotMean{\HKSScaleTenMaxData}{hksMaxPlot}
\addlegendentry{\HKSPlotMaxLegend}

\nextgroupplot[hksPanelB]
\HKSPlotMean{\HKSScaleThirtyFiveMedianData}{hksMedianPlot}
\HKSPlotMean{\HKSScaleThirtyFiveFiveData}{hksFivePlot}
\HKSPlotMean{\HKSScaleThirtyFiveTwentyFiveData}{hksTwentyFivePlot}
\HKSPlotMean{\HKSScaleThirtyFiveMaxData}{hksMaxPlot}

\nextgroupplot[hksPanelC]
\HKSPlotMean{\HKSRankData}{hksRankPlot}
\node[font=\HKSPlotNoteFont,anchor=south east] at (rel axis cs:1,1)
  {\HKSPlotNoteC};

\nextgroupplot[hksPanelD]
\HKSPlotMean{\HKSDensityMedianData}{hksMedianPlot}
\HKSPlotMean{\HKSDensityFiveData}{hksFivePlot}
\HKSPlotMean{\HKSDensityTwentyFiveData}{hksTwentyFivePlot}
\HKSPlotMean{\HKSDensityMaxData}{hksMaxPlot}
\node[font=\HKSPlotNoteFont,anchor=south east] at (rel axis cs:1,1)
  {\HKSPlotNoteD};
\end{groupplot}
\end{tikzpicture}
\par\smallskip
\pgfplotslegendfromname{hks-discussion-legend}
\caption{Mean normalized convergence time $T/n$.
(a,b) Size sweeps; only the horizontal axis is logarithmic in (a), both axes
in (b). (c) Rank sweep at $\rho=10$; (d) density sweep, both at
$n=10{,}000$. Panel (d) has a logarithmic vertical axis.
The color legend applies to (a,b,d); (c) shows $k$-large with varying $k$.}
\label{fig:hks-discussion-runtime}
\end{figure}
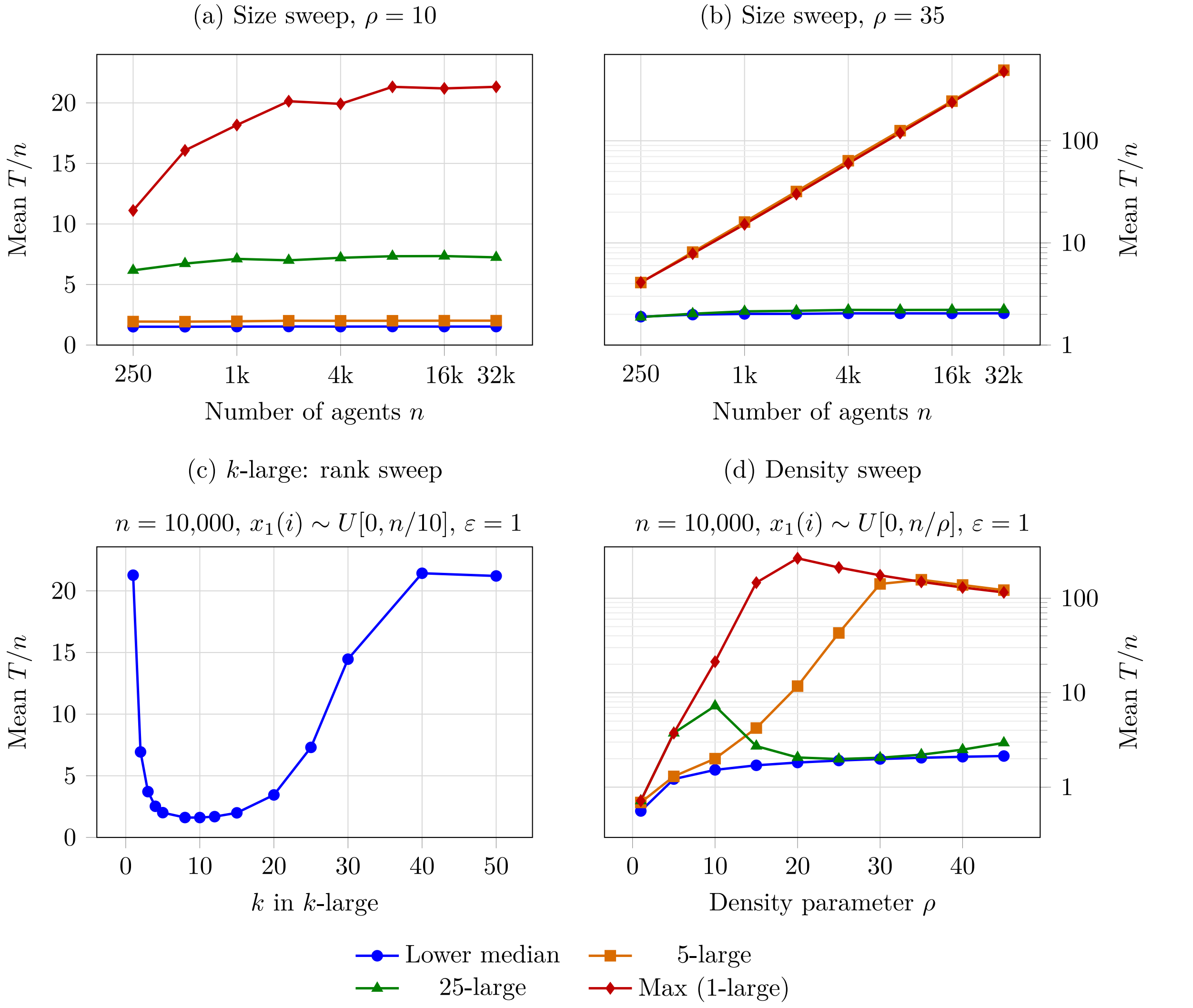
\endgroup

\Cref{fig:hks-discussion-runtime} (a,b) shows the convergence times for lower median, \largeagg{5}, \largeagg{25}, and \maxagg aggregation. For the median, $T/n$ is approximately constant at both displayed densities, as in the wider tested density range. At $\rho = 10$, the other rules also appear nearly linear at larger sizes. At $\rho = 35$, however, \maxagg and \largeagg{5} display near-quadratic growth: their fitted exponents over $n\in\{8000,16000,32000\}$ are approximately $2.00$ and $1.98$, respectively.

Using uniform random initialization over a closed interval, $\ell=n$ almost surely. Thus, our \medianagg and \largeagg{k} bounds become $\O(n^3)$. For \maxagg, we observed the sharper bound $\O(n^2)$, which matches its observed quadratic growth. Notably, \medianagg is much faster in our simulations, despite its weaker worst-case guarantee. This suggests that a sharper guarantee on the average-case convergence time might be possible. 

Keeping the initial profiles fixed while varying $k$ reveals a pronounced runtime minimum around $k=8$ to $k=10$ at $\rho=10$ (see (c)). Density also has non-monotone effects: \maxagg and \largeagg{5} peak near $\rho=20$ and $35$, then become faster, whereas \largeagg{25} peaks near $\rho=10$ (see (d)). Neighborhood masses and connectivity may jointly explain these runtime minima and density-dependent peaks. Further theoretical work could relate convergence time to structural properties of the instance and characterize the regimes of density and aggregation rank that yield linear or quadratic average-case convergence.

\subsection*{AI Disclosure Statement}

We posted an earlier version of this paper into ChatGPT 5.6 Sol (the tool). In the earlier version, we proved the existence of a potential function and a bound of $O(n^3)$ on the expected convergence time for asynchronous \largeagg{k}-HKS. The tool refined the proof to establish a worst-case bound of $O(n \ell^2)$. Moreover, it returned a proof of the worst-case convergence time of $O(n^2\ell^2)$ for asynchronous dynamics in 1-Lipschitz HKS. We significantly restructured and rewrote both proofs to obtain the versions presented here. 

We used the tool to implement the asynchronous dynamics and run the preliminary simulations presented in the discussion section. The setup and the goals of the experiments were defined by the authors. The code, data, and plots in the figure were generated by the tool.

The authors verified all results and wrote the entire text. We take full responsibility for the final work.

\subsection*{Acknowledgements}
This work was supported by DFG Research Unit ADYN (project number 411362735).

\bibliographystyle{abbrv}
\bibliography{references}

\end{document}